\documentclass[12pt]{iopart}

\usepackage{iopams}  
\usepackage{graphicx}
\usepackage{xcolor}
\usepackage{upgreek}
\begin{document}

\title[Energy fluctuations of a Brownian particle freely moving in a liquid]{Energy fluctuations of a Brownian particle freely moving in a liquid}

\author{Juan Ruben Gomez-Solano$^{1,*}$}

\address{$^{1}$Instituto de F\'isica, Universidad Nacional Aut\'onoma de M\'exico, Ciudad de M\'exico, C\'odigo Postal 04510, Mexico,}

\ead{$^*$ r\_gomez@fisica.unam.mx}
\vspace{10pt}
\begin{indented}
\item[]February 2024
\end{indented}

\begin{abstract}
We study the statistical properties of the variation of the kinetic energy of a spherical Brownian particle that freely moves in an incompressible fluid at constant temperature. Based on the underdamped version of the generalized Langevin equation that includes the inertia of both the particle and the displaced fluid, we derive an analytical expression for the probability density function of such a kinetic energy variation during an arbitrary time interval, which exactly amounts to the energy exchanged with the fluid in absence of external forces. We also determine all the moments of this probability distribution, which can be fully expressed in terms of a function that is proportional to the velocity autocorrelation function of the particle. The derived expressions are verified by means of numerical simulations of the stochastic motion of a particle in a viscous liquid with hydrodynamic backflow for representative values of the time-scales of the system. Furthermore, we also investigate the effect of viscoelasticity on the statistics of the kinetic energy variation of the particle, which reveals the existence of three distinct regimes of the energy exchange process depending on the values of the viscoelastic parameters of the fluid. 
\end{abstract}

%
\vspace{2pc}
\noindent{\it Keywords}: underdamped Brownian motion, heat fluctuations, stochastic thermodynamics, viscoelastic fluids, non-Markovian dynamics, Basset-Boussinesq force

%
%
%

\section{Introduction}\label{sect:intro}

Brownian motion is a hallmark of mesoscopic systems in contact with thermostats, which stems from their interaction with the surrounding molecules of the fluid media where they are commonly immersed, thereby involving intricate energy exchange processes subject to thermal fluctuations, such as those occurring in  molecular motors, biopolymers, colloidal particles, microorganisms and micron-sized devices \cite{Ciliberto2017}. Over the last few decades, the research on such small-scale energy transfers has been largely advanced thanks to the development of stochastic thermodynamics, which provides a conceptual framework to define quantities like internal energy, work, heat and entropy production at the level of single stochastic trajectories for thermostatted systems driven arbitrarily far from thermal equilibrium \cite{Sekimoto1998,Seifert2012,VANDENBROECK20156}. Based on stochastic thermodynamics, fundamental relations that constrain general properties of the probability distributions of the abovementioned quantities have been derived \cite{Jarzynski1997,Crooks1998,Hatano2001,Evans2002,Seifert2005}, e.g. the asymmetry in the probability of observing rare events of total entropy consumption with respect to entropy production for systems in non-equilibrium steady states, thus representing extensions of the second law of classical  thermodynamics to the mesoscopic realm.

Besides the general bounds imposed by such fluctuation relations, the study of the detailed statistics of stochastic thermodynamic quantities is relevant in situations where the energy exchange process takes place over time intervals that are comparable to the characteristic time-scales of a Brownian system in contact with a thermal bath, during which fluctuations are expected to exceed the mean value of the quantity of interest. The statistical properties of the heat transferred between the system and its surroundings are of particular relevance owing to the fact that energy is thermally injected into the system and dissipated into the bath even in the absence of external time-dependent forces that could apply work. Indeed, numerous  investigations have aimed at finding the probability density function of the heat transferred between an overdamped Brownian particle under a confining potential and coupled to a Markovian bath, as it is a good approximation for the dynamics of many experimental systems embedded in aqueous environments. For instance, the probability density function of the heat exchanged with a single bath at constant temperature has been investigated by means of the solution of Fokker-Planck-like equations \cite{Speck2005,Imparato2007}, path integrals \cite{Chatterjee2010,Chatterjee_2011}, and characteristic functions \cite{Fogedby_2009,Paraguassu_2021,PARAGUASSU2022126576}. In addition, by use of such methods the statistics of heat fluxes in overdamped systems under more complex conditions have been studied, such as confined Brownian particles and elastic manifolds in non-stationary states relaxing toward thermal equilibrium after a temperature quench \cite{GomezSolano2011,GomezSolano2012,Crisanti2017,Wu2023}, confined particles in presence of external flow \cite{Pagare2019,Saha2024}, and systems in contact with two thermostats at different temperatures, e.g. RC electric circuits \cite{Ciliberto2013,Ghosal2016}, and pairs of hydrodynamically-coupled colloidal particles \cite{Berut2016}. Moreover, the probability density function of the exchanged heat has also been analyzed for overdamped systems in contact with non-Markovian environments,  such as for Brownian particles harmonically trapped in active media  \cite{Goswami2019} and viscoelastic fluids \cite{Chatterjee2009,Darabi2023}. On the other hand, even though inertial terms have been long included in the description of energy fluctuations mainly in mesoscopic models of heat conduction \cite{Fogedby2011,Saito2011,Kundu_2011,Fogedby_2012,DHAR201549}, it was only recently recognized that kinetic energy variations of a system associated to its momentum degrees of freedom cannot be generally neglected in the statistical description of heat fluctuations \cite{Paraguassu2022}. Otherwise, this could result in inconsistencies as those clearly demonstrated in model systems such as a Brownian harmonic oscillator under sudden changes of the bath temperature \cite{Arold2018}, a Brownian particle subject to a non-conservative force in contact with multiple reservoirs at different temperatures \cite{Murashita2016}, and even for the basic case of a Brownian particle freely moving without the action of potentials \cite{Paraguassu2022}. This has recently led to an increased interest in exploring models that take into account the full underdamped description for the heat exchange process of  Brownian systems either at equilibrium and under non-equilibrium transformations even in instances where the overdamped approximation for the dynamics is expected to be valid at sufficiently long time-scales \cite{Rosinberg_2016,Rosinberg2017,Nicolis2017,Kwon2018,Fogedby_2020,Colmenares2022,PARAGUASSU2023128568,PARAGUASSU2023,Paraguassu2024}. 

It should be pointed out that in most of the simplified models employed to analyze mesoscopic energy transfers including inertia, memory effects that naturally emerge in the dynamics of actual experimental systems due its coupling with the environment are usually disregarded. Of special significance for Brownian motion at short time-scales in a liquid is the Basset-Boussinesq force, which originates from the unsteady flow of the entrained liquid around an accelerated particle and becomes important when the liquid density is comparable to that of the particle \cite{Stokes1851,Boussinesq1885,Basset1888,SCHIEBER20133}. Indeed, a particle transfers momentum to the fluid as it accelerates through it, thus forming vortices in its vicinity that gradually diffuse outwards, which affect the particle motion at later times if the particle translates over its own radius before the fluid relaxes. It should be noted that in most of the experimental conditions of colloidal suspensions dispersed in a liquid, the characteristic vortex-diffusion time is comparable to the momentum relaxation time of the suspended particles, and therefore the Basset-Boussinesq force cannot be neglected at such short inertial time-scales \cite{kim2005microhydrodynamics}. Moreover, as observed in pioneering molecular-dynamics simulations \cite{Rahman1964,Alder1967}, and later confirmed in experiments \cite{Paul1981,Huang2011,Jannasch2011,Kheifets2014,Mo2019}, the hydrodynamic memory caused by the Basset-Boussinesq force gives rise to a long-time power-law tail with exponent $-3/2$ in the decay of the velocity autocorrelation function (VACF) of a particle moving in a liquid  \cite{MR0014645,Zwanzig1970}, which greatly differs from the pure exponential relaxation predicted by the standard Langevin theory \cite{Langevin1908}, the latter being more suitable to describe Brownian motion in a dilute gas \cite{Mo2019,Li2010}. Another type of memory effect that commonly occurs in the dynamics of mesoscopic particles immersed in complex fluids, e.g. polymeric fluids and dense colloidal suspensions, arises from the viscoelasticity of the medium that can be described by a frequency-dependent complex viscosity accounting for both energy storage and dissipation \cite{bird1987dynamics}. In such a case, the slow relaxations of the fluid due to its crowded macromolecular microstructure gives rise to retarded effects in the drag force acting on the particle, which leads to transient subdiffusive behavior on time-scales that can be orders of magnitude larger than the inertial ones  \cite{Zwanzig1970,Mason1997,Squires2010}. At thermal equilibrium, both hydrodynamic and viscoelastic memory effects can be directly incorporated in the generalized Langevin equation \cite{Kubo1966,Zwanzig1973}, which represents the basis for the study of diverse aspects of Brownian systems occurring on inertial time-scales including hydrodynamic backflow, such as the single-particle motion in complex fluids \cite{Felderhof2009,Cordoba2012,PhysRevE.85.021504,PhysRevE.88.040701,PhysRevE.89.012130,Makris2021}, stochastic resonance \cite{He2020}, nonequilibrium fluctuation relations \cite{GASPARD2020121823}, transport phenomena \cite{Goychuk2019,Seyler2019,Goychuk2020,Seyler2020,Jiao2023}, and barrier crossing dynamics \cite{Cherayil2022}. Therefore, it is natural to use this framework in order to investigate the statistics of heat fluctuations of small systems taking place in liquid environments under more realistic conditions.

The main goal of the present paper is to investigate the effect of non-Markovianity due to hydrodynamic backflow and viscoelasticity on the probability distribution of the energy randomly exchanged as heat by a spherical Brownian particle with its surrounding liquid medium at constant temperature in absence of externally applied forces. To this end, by use of the generalized Langevin equation that includes the Basset-Boussinesq force, the rheological properties of the medium and additive thermal forces with the appropriate temporal correlations, we derive a general analytic expression for the probability density function of the energy transferred between the particle and the liquid over a time interval of arbitrary duration, whose even moments can be explicitly expressed in terms of the particle VACF. In the case of a Newtonian liquid of constant viscosity, our results are verified by means of numerical simulations of the non-Markovian dynamics of the particle including hydrodynamic memory. Then, using a Maxwell-like model for the stress relaxation modulus of the liquid, in addition to the Basset-Boussinesq force we also investigate the effect of viscoelasticity on the statistics of the heat, thereby uncovering the existence of three different regimes of the energy exchange process depending on the stress relaxation time and the viscosity ratio of the liquid.

The paper is organized as follows. In Section \ref{sect:model} we describe the model for the motion of a Brownian particle in a liquid at constant temperature, and identify some useful relations that characterize its dynamics in absence of externally applied forces. In Section \ref{sect:thermo} we derive expressions for the probability density function of the energy exchanged over an arbitrary time interval, as well as for all its moments, and discuss its relation with previously determined formulae for underdamped and overdamped Brownian systems without either hydrodynamic or viscoelastic memory effects induced by the medium. In Section \ref{sect:results} we present numerical results that illustrate the behavior of the different statistical quantities characterizing the heat fluctuations of a particle moving in a purely viscous liquid for representative values of the parameters of the system that could occur in experimental situations. In the same Section, we also analyze the influence of distinct viscoelastic parameters of the liquid on the statistics of the energy exchange of the embedded particle. Finally, in Section \ref{sect:summ} we summarize the main results of the paper and conclude.

\section{Model}\label{sect:model}

We are interested in the dynamics and energetics of  a micron- or submicron-sized spherical particle of mass density $\rho$ and radius $a$ moving with no-slip boundary conditions in an incompressible fluid, e.g. a liquid, of constant mass density $\rho_{\mathrm{f}}$, which is kept at constant temperature $T$, thereby acting as a thermal reservoir for the bead. The linear rheological properties of the fluid are fully characterized by its stress relaxation modulus, $G(t)$, which is a slowly decaying function of time $t \ge 0$ such that $G(t \rightarrow \infty) = 0$ and $G(t) = 0$ for $t < 0$ by causality \cite{bird1987dynamics}. For the sake of simplicity, we neglect the presence of gravity and focus on a single coordinate of the center of mass of the bead, $x$, whose corresponding velocity $v = \dot{x}$ evolves stochastically in time according to the generalized Langevin equation~\cite{Kubo1966}
\begin{equation}\label{eq:GLE}
   M \dot{v}(t) = -\int_0^t dt' \, \Gamma(t-t') v(t')  + \zeta(t).
\end{equation}
In Eq.~(\ref{eq:GLE}), the term on the left-hand side represents the net force acting on the particle, which causes its instantaneous acceleration $\dot{v}(t)$ at time $t\ge 0$ starting from the initial conditions $x_0 = x(0)$ and $v_0 = v(0)$ at $t = 0$. Here, $M$ is the effective mass of the particle, which is given by
\begin{equation}\label{eq:mass}
	M = m + \frac{1}{2} m_{\mathrm{f}}, 
\end{equation}
and includes the particle's own mass, $m = \frac{4}{3} \pi a^3 \rho $, plus a virtual mass equal to half of the mass of the fluid displaced by the particle, $m_{\mathrm{f}} = \frac{4}{3} \pi a^3 \rho_{\mathrm{f}}$, which accounts for the inertia of the latter~\cite{Stokes1851,Boussinesq1885,Basset1888,SCHIEBER20133}. Moreover, the first term on the right-hand side of Eq.~(\ref{eq:GLE}) corresponds to the coarse-grained force exerted on the bead at time $t$ by its interaction with all the surrounding fluid particles, whose delayed effect at times $0 \le t' \le t$ is encoded by the memory kernel $\Gamma(t-t')$,  with $\Gamma(t - t') = 0$ for $t' > t$ by causality. This function encompasses the non-Markovianity of the particle motion induced by both the rheological properties of the fluid and the hydrodynamic backflow produced by the entrained fluid. In the case of no-slip boundary conditions on the particle surface and at low Reynolds number, the Fourier transform of the memory kernel, $\hat{\Gamma}(\omega) = \int_{-\infty}^{\infty} dt\, e^{-i\omega t} \Gamma(t) =  \int_0^{\infty} dt\, e^{-i\omega t} \Gamma(t)$, is explicitly given by \cite{Zwanzig1970,PhysRevE.85.021504,XU2007150}
\begin{equation}\label{eq:GSFour}
    \hat{\Gamma}(\omega) = 6\pi a \hat{\eta}(\omega) + 6\pi a^2 \sqrt{i\omega \rho_{\mathrm{f}} \hat{\eta}(\omega)},
\end{equation}
where $i = \sqrt{-1}$, $\omega$ is a real-valued frequency and $\hat{\eta}(\omega)$ is a complex viscosity that depends on $\omega$, which is given by the Fourier transform of the fluid's relaxation modulus, $\hat{\eta}(\omega) = \int_{-\infty}^{\infty} dt \, e^{-i\omega t} G(t) = \int_0^{\infty} dt \, e^{-i\omega t} G(t)$ \cite{bird1987dynamics}. The convolution of the inverse Fourier transform of the first term on the right-hand side of Eq.~(\ref{eq:GSFour}) with the particle velocity leads to the generalized Stokes law for the drag force on a sphere moving in a fluid, $-6 \pi a \int_{0}^{t} dt' \, G(t - t') v(t') $, which reduces to the well-known expression $-6 \pi a \eta v(t)$ for viscous friction in the case of a purely viscous (Newtonian) fluid of constant viscosity $\eta$, whose corresponding relaxation modulus is $G(t) = 2\eta \delta(t)$ with $\delta(t)$ is the Dirac delta function. In addition, the inverse Fourier transform of the second term on the right-hand side of Eq.~(\ref{eq:GSFour}) represents the hydrodynamic part of  the memory kernel  that originates from the delayed interaction of the accelerated bead with the diffusing vortices generated in the fluid around it. Its convolution with the particle velocity gives rise to the Basset-Boussinesq force, which does not have a simple general expression for an arbitrary complex viscosity of the fluid. In the case of a Newtonian fluid, i.e. constant $\hat{\eta}(\omega) = \eta$, the Basset-Boussinesq force is given by $-6 a^2  \sqrt{\pi \rho_{\mathrm{f}} \eta}  \int_0^t  dt' \, (t - t')^{-1/2} \dot{v}(t') $, thus reflecting the long-ranged memory induced by the entrained fluid on the particle motion. Furthermore, the term $\zeta(t)$ in Eq.~(\ref{eq:GLE})  is a Gaussian colored noise that results from the thermal collisions of the particles that make up the fluid, whose mean and autocorrelation function satisfy for all $t \ge 0$ \cite{kubo2012statistical}
\begin{eqnarray}\label{eq:FD2nd}
    \langle \zeta(t) \rangle & = & 0, \nonumber\\
    \langle \zeta(t) \zeta(0) \rangle & = & k_B T \Gamma(t),
\end{eqnarray}
respectively.  Note that,  for the sake of convenience, we have set the lower limit of integration to $t' = 0$ in Eq.~(\ref{eq:GLE}) in order to carry out the calculations presented in the rest of this Section  in a straightforward manner. This corresponds to an apparent non-stationary situation where the Gaussian noise $\zeta(t)$ has an autocorrelation function that depends on the arbitrarily chosen time origin at $t' = 0$, as expressed by Eq.~(\ref{eq:FD2nd}). However, Eq.~(\ref{eq:GLE}) represents exactly the same stochastic dynamics as that described by the following generalized Langevin equation with time-translation invariance for all $t > -\infty$
\begin{equation}\label{eq:GLE2}
	M \dot{v}(t) = -\int_{-\infty}^t dt' \, \Gamma(t-t') v(t')  + \xi(t),
\end{equation}
provided that $\langle v(0) \zeta(t)\rangle = 0$ for all $t > 0$~\cite{kubo2012statistical}. In Eq.~(\ref{eq:GLE2}), $\xi(t)$ is a Gaussian process with the following autocorrelation function at thermal equilibrium
\begin{equation}\label{eq:zetaxi}
 	\langle \xi(t') \xi(t' + t) \rangle  =  k_B T \Gamma(t) 
\end{equation}
for all $t' > -\infty$ and $t \ge 0$, i.e. it is stationary, which requires that the Gaussian process $\zeta(t)$ in Eq.~(\ref{eq:GLE}) be
\begin{equation}\label{eq:noisexi}
	\zeta(t) = \xi(t) - \int_{-\infty}^0 dt' \, \Gamma(t-t') v(t'),
\end{equation}
such that $ \langle \xi(t') \xi(t' + t) \rangle  = \langle \zeta(t) \zeta(0) \rangle$.

We now focus on the particle velocity $v(t)$ at time $t > 0$ that evolves according to Eq. (\ref{eq:GLE}) starting from the initial condition $v(0) = v_0$ and a given stochastic realization of the thermal noise $\{ \zeta(t'), 0 \le t' \le t \}$. By computing the Laplace transform of each term in  Eq. (\ref{eq:GLE}), we find that the Laplace transform of $v(t)$, $\tilde{v}(s) = \int_0^{\infty} dt \, e^{-st} v(t)$, can be expressed as
\begin{equation}\label{eq:Lapsolvel}
	\tilde{v}(s) = v_0 \tilde{\psi}(s) + \frac{1}{M} \tilde{\psi}(s) \tilde{\zeta}(s),
\end{equation}
where the function $\tilde{\psi}(s)$ depends on the complex-valued frequency $s$ and is given by the expression
\begin{equation}\label{eq:Lappsi}
	\tilde{\psi}(s) = \frac{1}{s + \frac{1}{M} \tilde{\Gamma}(s)},
\end{equation}
with $\tilde{\Gamma}(s) = \int_0^{\infty} dt \, e^{-st} \Gamma(t)$ the Laplace transform of the memory kernel $\Gamma(t)$. By inverting Eq.~(\ref{eq:Lapsolvel}), we find the solution $v(t)$ in time domain
\begin{equation}\label{eq:solvel}
	v(t) = v_0 \psi(t) + \frac{1}{M} \int_0^t dt' \, \psi(t -t') \zeta(t'),
\end{equation}
where $\psi(t)$ is the inverse Laplace transform of the function defined in Eq.~(\ref{eq:Lappsi}). We are interested in a situation where the bead is in equilibrium with the surrounding liquid, with no external forces applied to the system. Therefore, for all times $t \ge 0$ the probability density function of the particle velocity $v$ must correspond to the Boltzmann distribution
\begin{equation}\label{eq:BD}
	P_{eq}(v) = \sqrt{\frac{M}{2\pi k_B T}} \exp \left( - \frac{M v^2}{2k_B T} \right),
\end{equation}
which involves the effective mass $M$ instead of $m$ in the definition of the kinetic energy of the particle, $\frac{1}{2}Mv^2$, because of the contribution of the virtual mass added by the inertia of the entrained fluid, as experimentally verified for microspheres trapped in water and acetone \cite{Mo2015}. This is consistent with a modified equipartition relation for $v$ at all times $t \ge 0$
\begin{eqnarray}\label{eq:MEP}
	\frac{1}{2}M \langle v^2 \rangle_{eq} & = & \frac{1}{2}M\int_{-\infty}^{\infty} dv \, P_{eq}(v) v^2, \nonumber\\ 
		& = & \frac{1}{2} k_B T, 
\end{eqnarray}
where $\langle \ldots \rangle_{eq}$ denotes an average with respect to the equilibrium velocity distribution $P_{eq}(v)$. Then, by multiplying both sides of Eq.~(\ref{eq:solvel}) by $v_0$ and then computing the average with respect to the Boltzmann distribution (\ref{eq:BD}), it follows that the VACF at equilibrium is related to the inverse Laplace transform of the function $\psi(t)$ defined in Eq. (\ref{eq:Lappsi}) by 
\begin{equation}\label{eq:VACF}
	\langle v(\tau) v(0) \rangle_{eq} = \frac{k_B T}{M} \psi(\tau).
\end{equation}
where $\tau > 0$ represents the time elapsed between the initial and the final velocities. It is important to realize that, regardless of the specific form of the memory kernel $\Gamma(t-t')$, the function $\psi(\tau)$ satisfies
\begin{eqnarray}\label{eq:psilim}
        \lim_{\tau \rightarrow 0} \psi(\tau) &= \lim_{s\rightarrow \infty} s\tilde{\psi}(s) = 1,\nonumber\\
        \lim_{\tau \rightarrow \infty} \psi(\tau) &= \lim_{s\rightarrow 0} s\tilde{\psi}(s) = 0.
\end{eqnarray}
Therefore, the VACF in Eq. (\ref{eq:VACF}) must decay to zero in the long-time limit, whereas it approaches the value of the variance given by the modified equipartition relation (\ref{eq:MEP}), $\langle v(\tau) v(0)\rangle_{eq} \rightarrow \frac{k_B T}{M}$, as $\tau \rightarrow 0$. 

Following the same approach as in \cite{Darabi2023}, we proceed to derive some expressions that will be useful for the statistics of the energy exchanged between the particle and the liquid. First, by computing the average of $\left[ v(\tau) - v_0 \psi(\tau) \right]^2$ over an infinite number of realizations of the thermal noise  $\{ \zeta(t), 0 \le t \le \tau \}$ for a fixed value of the initial velocity $v_0$, and then with respect to the Boltzmann distribution in Eq. (\ref{eq:BD}) for an ensemble of equilibrium values of $v_0$, by use of Eq. (\ref{eq:VACF}) we find that
\begin{equation}\label{eq:sqvel}
	\frac{k_B T}{M} \left[ 1 - \psi(\tau)^2 \right] = \frac{1}{M^2} \int_0^{\tau} dt \int_0^{\tau} dt' \left\langle \zeta(t) \zeta(t') \right\rangle \psi(\tau - t) \psi(\tau -  t'),
\end{equation}
where $\langle \ldots \rangle$ represents the average only with respect to the noise at fixed $v_0$. Eq. (\ref{eq:sqvel}) facilitates the determination of an expression for the conditional probability density of finding the particle with velocity $v$ at time $t = \tau>0$, provided that its velocity was $v_0$ at time $t =0$, which we denote as $P(v,\tau|v_0,0)$. Indeed, by use of the Fourier representation of the Dirac delta function, $\delta \left[ v - v(\tau) \right] = \frac{1}{2\pi} \int_{-\infty}^{\infty} dk\, e^{ik \left[ v - v(\tau) \right]}$, as well as the time-domain solution $v(t)$ at time $t = \tau > 0$ with initial condition $v(0) = v_0$ for the particle velocity shown in Eq. (\ref{eq:solvel}), such a conditional probability density can be expressed as
\begin{eqnarray}\label{eq:condpdfv}
    \fl	P(v,\tau|v_0,0) & = & \langle \delta\left[ v  - v(\tau)  \right] \rangle,\nonumber\\
	\fl		& = & \frac{1}{2\pi} \int_{-\infty}^{\infty} dk \, e^{  ik \left[ v - v_0\psi(\tau) \right]  } \left \langle \exp \left[ -ik \frac{1}{M}\int_0^{\tau} dt \, \psi (\tau-t) \zeta(t) \right] \right \rangle.
\end{eqnarray}
It should be noted that the term $\Phi \left[-k  \psi / M \right] \equiv \left \langle \exp \left\{ i \int_0^{\tau} dt \, \left[- \frac{k}{M} \psi (\tau-t) \right] \zeta(t)  \right\} \right \rangle$ in the second line of Eq. (\ref{eq:condpdfv}) is nothing else than the characteristic functional of the thermal noise $\zeta(t)$ for the function $-k\psi(\tau -t)/M$ over the time interval $0 \le t \le \tau$. Since we assume that $\zeta(t)$ is Gaussian, by virtue of Eq. (\ref{eq:sqvel}) the characteristic functional $\Phi[-k \psi / M]$ can be simply expressed as~\cite{kubo2012statistical}
\begin{eqnarray}\label{eq:charactfunc}
	\Phi[- k\psi /M] & = & \exp \left[ -\frac{k^2}{2 M^2} \int_0^{\tau} dt \int_0^{\tau} dt'  \left\langle \zeta(t) \zeta(t') \right\rangle \psi(\tau-t) \psi(\tau-t')  \right], \nonumber\\
			& = & \exp \left\{ - k^2 \frac{k_B T}{2 M}\left[ 1 - \psi(\tau)^2 \right] \right\}.
\end{eqnarray}
Finally, a straightforward calculation of the integral in Eq. (\ref{eq:condpdfv}) over the wavenumber $-\infty < k < \infty$ using the explicit formula of Eq. (\ref{eq:charactfunc}) for the characteristic functional of the thermal noise leads the following expression for the conditional probability density $P(v,\tau|v_0,0)$
\begin{equation}\label{eq:pdfcondvel}
    P(v,\tau|v_0,0) = \sqrt{\frac{M}{2\pi k_B T\left[1-\psi(\tau)^2 \right]}} \exp \left\{ -\frac{M}{2k_BT} \frac{\left[ v - v_0 \psi(\tau) \right]^2}{1 - \psi(\tau)^2}\right\}.
\end{equation}
Noteworthy, the assumption that the initial velocity $v_0$ at time $t = 0$ is distributed according to the equilibrium Boltzmann distribution guarantees that the probability density function of $v$ remains the same at all times $\tau > 0$ in absence of external forces, i.e. 
\begin{equation}\label{eq:statpdfvel}
	\int_{-\infty}^{\infty} dv_0 \, P(v,t|v_0,0) P_{eq}(v_0) = P_{eq}(v), 
\end{equation}
as can be verified by direct substitution of Eq. (\ref{eq:pdfcondvel}) into Eq. (\ref{eq:statpdfvel}) with both $P_{eq}(v_0)$ and $P_{eq}(v)$ given by Eq. (\ref{eq:BD}).

Some remarks must be made regarding Eqs. (\ref{eq:BD}), (\ref{eq:MEP}) and (\ref{eq:VACF}). There is seemingly a discrepancy between the variance of the particle velocity given by Eq. (\ref{eq:MEP}), $k_B T/M$, and the larger value provided by the equipartition theorem for the kinetic energy of the particle alone, $ k_B T / m$. However, the apparent conflict is due to the assumption of perfect incompressibility of the fluid, which breaks down even in liquids for particle velocities measured on time-scales that are shorter than the time needed by a sound wave to propagate over a distance equal to the particle radius, $\tau_s = a/c$, where $c$ is the speed of sound in the liquid~\cite{MR0014645,Zwanzig_Bixon_1975,Giterman1966}. During time intervals that are much smaller than $\tau_s$, the compressibility of the liquid cannot be neglected and the particle motion actually decouples from the entrained fluid, thus leading to the conventional equipartition relation $m \langle v^2 \rangle_{eq} / 2 = k_B T/2 $ \cite{Giterman1966}. This effect translates into an actual drop of the VACF from ${k_B T}/{m}$ to ${k_B T}/{M}$ over the time interval $0 \le \tau \lesssim \tau_s$ due to acoustic damping of the particle velocity \cite{Zwanzig_Bixon_1975}. Therefore, the behavior of the VACF described by Eq. (\ref{eq:VACF}) with $\psi(\tau)$ defined in Eq. (\ref{eq:Lappsi}) is strictly valid only for time intervals $\tau \gtrsim \tau_s$, over which the liquid can be considered as incompressible. It is worth mentioning that the acoustic propagation time $\tau_s$, whose values are typically a fraction of nanoseconds for, e.g. micron-sized spheres in water, is several orders of magnitude smaller than the corresponding microsecond values of the inertial time-scales of the particle and the liquid, and therefore Eqs. (\ref{eq:BD}), (\ref{eq:MEP}) and (\ref{eq:VACF}) describe correctly the statistics of the particle velocity measured on time intervals larger than the former, as verified in experiments \cite{Kheifets2014,Mo2015}.  For the level of description required in the present paper, it will be enough to assume that the fluid is fully incompressible so that the variance of the particle position, which corresponds to the limit of the VACF when $\tau \rightarrow 0$, is simply given by the the modified equipartition relation (\ref{eq:MEP}), whereas the kinetic energy of the particle must include the added mass of the displaced fluid.

\section{Statistics of the exchanged energy}\label{sect:thermo}
In this section, we derive an expression for the probability density function of the energy transferred from the liquid, which behaves as a thermal reservoir when kept at constant temperature, to the embedded particle. Since we are interested in a situation where there are no external forces performing work on the system, the energy exchanged between the particle and the liquid during a time interval $[0,\tau]$ at the level of a single stochastic trajectory of the particle starting at $[x(0) = x_0,v(0) = v_0]$ and ending at $[x(\tau) = x_{\tau}, v(\tau) = v_{\tau}]$, which we denote as $\mathcal{Q}_{\tau}$, can be calculated from the corresponding variation of its kinetic energy
\begin{equation}\label{eq:heat}
    \mathcal{Q}_{\tau}  = \frac{M}{2} \left( v_{\tau}^2 - v_0^2\right).
\end{equation}
If $\mathcal{Q}_{\tau} > 0$, the particle absorbs energy from the surroundings, whereas it releases energy if $\mathcal{Q}_{\tau} < 0$, thus increasing or decreasing its kinetic energy, respectively. Since $\mathcal{Q}_{\tau}$ is simply determined by the difference of the two stochastic variables $v_0^2$ and $v_{\tau}^2$ according to Eq.~(\ref{eq:heat}), which are in general correlated for finite values values of $\tau$, an expression for the probability density function of the energy exchanged during a time interval of duration $\tau > 0$ can be derived by means of the equation
\begin{equation}\label{eq:pdfqtau}
P(\mathcal{Q},\tau)  =  \int_{-\infty}^{\infty} dv_{\tau} \int_{-\infty}^{\infty} dv_0 P_{eq}(v_0) P(v_{\tau},\tau|v_0,0) \delta \left[ \mathcal{Q} - \frac{M}{2} \left( v_{\tau}^2 - v_0^2\right) \right].
\end{equation}
In Eq.~(\ref{eq:pdfqtau}) $P_{eq}(x_0)$ is the probability density function of the initial velocity $v_0$ given by (\ref{eq:BD}) since we assume that the system is at all times $t \ge 0$ in thermal equilibrium with the surrounding fluid, Furthermore, the function $P(v_{\tau},\tau|v_0,0)$ in~Eq.~(\ref{eq:pdfqtau}) represents the conditional probability density that the particle has a velocity $v_{\tau}$ at time $t = \tau > 0$ provided that at time $t = 0$ its velocity was $v_0$, whose explicit formula is given in Eq.~(\ref{eq:pdfcondvel}) for any arbitrary memory kernel $\Gamma(t-t')$ governing the stochastic dynamics of $v(t)$. Bearing in mind that the Dirac delta function in the integrand of Eq.~(\ref{eq:pdfqtau}) can be expressed as
\begin{equation}\label{eq:diracdeltaq}
    \delta \left[ \mathcal{Q} - \frac{M}{2} \left( v_{\tau}^2 - v_0^2\right) \right] = \frac{ \delta\left(v_{\tau} - \sqrt{v_0^2 + \frac{2}{M}\mathcal{Q}} \right) + \delta\left(v_{\tau} + \sqrt{v_0^2 + \frac{2}{M}\mathcal{Q}} \right)}{M \sqrt{v_0^2 + \frac{2}{M}\mathcal{Q}}},
\end{equation}
and then substituting~Eq.~(\ref{eq:diracdeltaq}) into Eq.~(\ref{eq:pdfqtau}), upon integration over $-\infty < x_{\tau} < \infty$ que probability density $P(\mathcal{Q},\tau)$ can be written as
\begin{equation}\label{eq:pdfqtau2}
P(\mathcal{Q},\tau) = P_+(\mathcal{Q},\tau) + P_-(\mathcal{Q},\tau),
\end{equation}
where $P_+(\mathcal{Q},\tau)$ and $P_-(\mathcal{Q},\tau)$ are defined as
\begin{equation}\label{eq:pdfpm}
    P_{\pm}(\mathcal{Q},\tau) = \frac{1}{M} \int_{-\infty}^{\infty} dv_0 \frac{P_{eq}(v_0) P\left(\pm \sqrt{v_0^2 + \frac{2}{M}\mathcal{Q}},\tau|v_0,0\right)}{\sqrt{v_0^2 + \frac{2}{M}\mathcal{Q}}}.
\end{equation}
In addition, the change of variables 
\begin{equation}\label{eq:changez}
    z_{\pm} = \sqrt{\frac{M}{k_B T}} \left( \sqrt{v_0^2 + \frac{2}{M} \mathcal{Q}} \mp v_0 \right),
\end{equation}
allows one to express the integrals in Eq.~(\ref{eq:pdfpm}) as
\begin{equation}\label{eq:pdfpm2}
    P_{\pm}(\mathcal{Q},\tau) = \frac{1}{2\pi k_B T \sqrt{\phi_+(\tau) \phi_-(\tau)}} \int_0^{\infty} dz_{\pm} \, \frac{1}{z_{\pm}} \exp\left[-\frac{\left(\frac{\mathcal{Q}}{k_B T}\right)^2}{\phi_{\pm}(\tau) z_{\pm}^2}  - \frac{z_{\pm}^2}{4 \phi_{\mp}(\tau) } \right],
\end{equation}
where the functions  $\phi_{+}(\tau)$ and $\phi_{-}(\tau)$ are defined as
\begin{equation}\label{eq:phipm}
    \phi_{\pm}(\tau) = 1 \pm \psi(\tau).
\end{equation}
Moreover, by use of the identity
$\int_0^{\infty} dz \, z^{\nu - 1} e^{ -\beta z^p - \alpha z^{-p} } = \frac{2}{p} \left( \frac{\alpha}{\beta} \right)^{\frac{\nu}{2p}} K_{\frac{\nu}{p}}\left( 2 \sqrt{\alpha \beta} \right)$, where $K_{\frac{\nu}{p}}(2\sqrt{\alpha \beta})$ is the ${\nu}/{p}-$th order modified Bessel function of the second kind with argument $2\sqrt{\alpha \beta}$ \cite{gradshteyn2014table}, it can be shown that $P_{+}(\mathcal{Q},\tau) = P_{-}(\mathcal{Q},\tau)$. Therefore, Eq.~(\ref{eq:pdfqtau}) simplifies to
\begin{equation}\label{eq:pdfQ}
P(\mathcal{Q},\tau)=\frac{1}{\pi k_B T\sqrt{1-\psi(\tau)^2}}K_0\left(\frac{1}{\sqrt{1-\psi(\tau)^2}}\frac{\vert\mathcal{Q}\vert}{k_BT}\right),
\end{equation}
where $K_0(w) = \int_0^{\infty} du \, \cos \left( w \sinh u  \right)$ denotes the zeroth order modified Bessel function of the second kind. Remarkably, despite the complexity of the non-Markovian particle dynamics in a liquid described by Eq. (\ref{eq:GLE}), the probability density function of the exchanged energy has a rather simple form, where all the information on the coupling between the particle and the its environment is encoded in the function $\psi(\tau)$, which is proportional to the particle VACF according to Eq.~(\ref{eq:VACF}).

From the expression for the probabilidty density $P(\mathcal{Q},\tau)$ given in Eq.~(\ref{eq:pdfQ}), we can determine all its moments of order $n \ge 0$ 
\begin{equation}\label{eq:qmoment}
    \langle \mathcal{Q}^n_{\tau} \rangle = \int_{-\infty}^{+\infty}d\mathcal{Q} \, P(\mathcal{Q},\tau) \mathcal{Q}^n, 
\end{equation}
which are explicitly given by the following expressions
\begin{equation}\label{eq:momentsq}
   \langle \mathcal{Q}^n_{\tau} \rangle = \left\{
    \begin{array}{ll}
   0, & \,\,\,\,\, \mathrm{if} \,n \, \mathrm{is}\, \mathrm{odd},\\
   \left[ \frac{n!}{\left( \frac{n}{2}\right)!}  \right]^2 \left( \frac{k_B T}{2} \right)^n \left[ 1 - \psi(\tau)^2\right]^{\frac{n}{2}}, & \,\,\,\,\,  \mathrm{if} \,n \, \mathrm{is}\, \mathrm{even}.
    \end{array} \right.
\end{equation}
In Eq.~(\ref{eq:momentsq}) we have made use of the symmetry of $P(\mathcal{Q},\tau)$ with respect to $\mathcal{Q} = 0$, as well as of the definite-integral identity $\int_0^{\infty} dw \, w^{\lambda-1}K_{\mu}(w) = 2^{\lambda -2} \mathit{\Gamma}(\frac{\lambda-\mu}{2}) \mathit{\Gamma}(\frac{\lambda+\mu}{2})$, where $\mathit{\Gamma}(z) = \int_0^{\infty} du \, u^{z-1} e^{-u}$ represents the gamma function, which must not be confused with the memory kernel $\Gamma(t-t')$ describing the particle dynamics in~Eq.~(\ref{eq:GLE}). Note that, by taking $n = 0$ in~Eq.~(\ref{eq:momentsq}), we can verify that $P(\mathcal{Q},\tau)$ is normalized, i.e. $\int_{-\infty}^{\infty} d\mathcal{Q} \, P(\mathcal{Q},\tau) = 1$. Moreover, for all values of the time interval $\tau > 0$, the mean energy transferred from the liquid to the particle is 
\begin{equation}\label{eq:meanq}
    \langle \mathcal{Q}_{\tau} \rangle = 0,
\end{equation}
which results from the fact that no net energy fluxes occur on average at thermal equilibrium. In addition, the variance of the exchanged energy during a time interval $\tau >0$, $ \langle \Delta \mathcal{Q}^2_{\tau}\rangle \equiv \langle \mathcal{Q}^2_{\tau} \rangle - \langle \mathcal{Q}_{\tau}\rangle^2$, is 
\begin{equation}\label{eq:varianceq}
   \langle \Delta \mathcal{Q}^2_{\tau}\rangle  = (k_B T)^2 \left[1 - \psi(\tau)^2 \right].
\end{equation}
According to the limits of the function $\psi(t)$ shown in Eq.~(\ref{eq:psilim}), the asymptotic value of the variance in Eq. (\ref{eq:varianceq}) as $\tau \rightarrow \infty$ is $\langle \Delta \mathcal{Q}^2_{\tau \rightarrow \infty}\rangle = (k_B T)^2$, irrespective of the specific features of the memory kernel $\Gamma(t-t')$ and the effective mass $M$ that determine the behavior of the function $\psi(\tau)$ through Eq. (\ref{eq:Lappsi}). This result shows that, for sufficiently long intervals over which all temporal correlations induced either by the Basset-Boussinesq force or by the viscoelasticity of the medium have already vanished, the characteristic exchanged energy is $\left( \langle \Delta \mathcal{Q}^2_{\tau \rightarrow \infty} \rangle\right)^{1/2} = \pm k_B T$.
Finally, from Eq.~(\ref{eq:momentsq}), we can also compute the skewness of $P(\mathcal{Q},\tau)$, which turns out to be 
\begin{equation}\label{eq:skew}
	\frac{ \langle {\mathcal{Q}}_{\tau}^3 \rangle - 3 \langle {\mathcal{Q}}_{\tau} \rangle \langle \Delta \mathcal{Q}^2_{\tau}\rangle - \langle \mathcal{Q}_{\tau} \rangle^3 }{\left( \langle \Delta \mathcal{Q}^2_{\tau}\rangle \right)^{3/2}}= 0, 
\end{equation}
whereas the kurtosis is independent of $\tau$ and has the constant value
\begin{equation}\label{eq:kurt}
    K_{\tau} \equiv \frac{\langle \mathcal{Q}^4_{\tau} \rangle}{ \langle \mathcal{Q}^2_{\tau} \rangle^2}  = 9.
\end{equation}

It is worth pointing out that Eq.~(\ref{eq:pdfQ}) represents a generalization of previosly derived expressions for the probability density function of the heat exchanged between a Brownian particle and its surroundings without external forces. Indeed, in absence of hydrodynamic backflow and viscoelasticity of the fluid, the memory kernel of a particle of mass  $m$ moving with friction coefficient $\gamma$ in a viscous medium at constant temperature $T$ is $\Gamma(t-t') = 2 \gamma \delta(t-t')$ for $t \ge t'$ and $\Gamma(t-t') = 0$ for $t < t'$, whereas its VACF is $\langle v(\tau) v(0) \rangle_{eq} = \frac{k_B T}{m} e^{-\frac{\gamma \tau }{m}}$, i.e. $\psi(\tau) = e^{-\frac{\gamma \tau }{m}}$ according to Eq.~(\ref{eq:VACF}). Therefore, in such a case 
\begin{equation}\label{eq:psiMarkov}
	 1 - \psi(\tau)^2 =  \frac{2} { 1 + \coth \left( \frac{\gamma \tau}{m} \right)}, 
\end{equation}
which yields 
\begin{equation}\label{eq:pdfQmarkov}
	P(\mathcal{Q},\tau) =  \frac{1}{\pi k_B T}   \sqrt{\frac{1}{2} \left[ 1+\coth \left( \frac{\gamma \tau}{m} \right) \right] }  K_0 \left(   \frac{|\mathcal{Q}|}{k_B T} \sqrt{\frac{1}{2} \left[ 1+\coth \left( \frac{\gamma \tau}{m} \right) \right] }  \right),
\end{equation}
and is exactly the same as the expressions derived in \cite{Paraguassu2022} and \cite{PARAGUASSU2023128568} using the characteristic function of the exchanged heat under the abovementioned memoryless conditions. In addition, regardless of the strength of the Basset-Boussinesq force and the viscoelastic features of the liquid, Eq. (\ref{eq:pdfQ}) reduces in the long-time limit $\tau \rightarrow \infty$ to the generic expression
\begin{equation}\label{eq:limitpdfQ}
	P(\mathcal{Q},\tau \rightarrow \infty) = \frac{1}{\pi k_B T} K_0 \left( \frac{|Q|}{k_B T} \right),
\end{equation}
which is also independent of the mass of the particle, and is identical to the expression derived in \cite{Nicolis2017} using the conventional Langevin equation without hydrodynamic and viscoelastic memory effects. Asymptotic expressions in terms of the zeroth order modified Bessel function of the second kind that are similar to Eq.~(\ref{eq:limitpdfQ}) have also been found for probability distributions of energy exchanges for particles confined by non-linear potentials at low temperature \cite{Fogedby_2009}, particles harmonically trapped in viscous baths in stationary \cite{Imparato2007,Chatterjee2010}, and non-stationary states \cite{GomezSolano2011,Crisanti2017}, as well as for particles moving in active media \cite{Goswami2019}. Interestingly, Eq. (\ref{eq:pdfQ}) has the same form as the probability density function of the variation of the potential energy of an overdamped Brownian particle confined by a harmonic trap in both viscous and viscoelastic fluids, where the function $\psi(\tau)$ is replaced by a function that is proportional to the autocorrelation function of the particle position~\cite{Chatterjee2010,Darabi2023}.

\section{Results}\label{sect:results}
In this Section, we analyze in detail the effect of hydrodynamic backflow and viscolasticity of the liquid in the stochastic energetics of an embedded Brownian bead in order to illustrate the general findings of Section \ref{sect:thermo}. To this end, it is convenient to identify the following characteristic scales of the system
\begin{eqnarray}
	\tau_c & = & \frac{M}{\gamma}, \label{eq:tauc}\\
	v_c & = & \sqrt{\frac{k_B T}{M}}, \label{eq:vc}\\
	E_c & = & k_B T, \label{eq:Ec}
\end{eqnarray}
which represent the effective momentum relaxation time of the particle including its virtual mass, the standard deviation of the fluctuations of the particle velocity, and the thermal energy from the bath, respectively. In Eq. (\ref{eq:tauc}), $\gamma = 6\pi a  \eta_{\infty}$ is a characteristic friction coefficient of the bead, with $\eta_{\infty} = \lim_{\omega \rightarrow \infty}  \hat{\eta}(\omega) > 0$ the high-frequency viscosity of the liquid due to the presence of a viscous solvent, where $\hat{\eta}(\omega) = \eta_{\infty}$ for all values of $\omega$ in the case of a Newtonian liquid. Additionally, the characteristic scales defined in Eqs.~(\ref{eq:tauc})-(\ref{eq:Ec}) allow us to define characteristic length and a force scales as follows
\begin{eqnarray}
	x_c & = & v_c \tau_c = \sqrt{\frac{Mk_B T}{\gamma^2}} , \label{eq:xc}\\
	F_c & = & \frac{E_c}{v_c \tau_c} = \sqrt{\frac{\gamma^2 k_B T}{M}}. \label{eq:Fc}
\end{eqnarray}
Non-dimensionalization of the different quantities describing the dynamics and energetics of the system by means of the characteristic scales defined in Eqs. (\ref{eq:tauc})-(\ref{eq:Fc}) permits the direct comparison of various scenarios representing possible experimental conditions. Likewise, it is helpful to introduce the following time-scale expressed in terms of the characteristic time $\tau_c$ defined in Eq.~(\ref{eq:tauc}) and the ratio of mass densities $\rho/\rho_{\mathrm{f}}$ as
\begin{equation}\label{eq:tauf}
	\tau_{\mathrm{f}}  =  \frac{\rho_{\mathrm{f}}a^2}{\eta_{\infty}} = \frac{9}{2\frac{\rho}{\rho_{\mathrm{f}}} + 1} \tau_c, 
\end{equation}
which stands for the time that a vortex formed around the particle due to the entrainment of the liquid needs to diffuse over a distance equal to the particle radius. where the corresponding diffusivity is $\eta_{\infty}/\rho_{\mathrm{f}}$. Note that the values of this vortex-diffusion time are bounded on the interval $0 \le \tau_{\mathrm{f}} \le 9\tau_c$, where the lower and upper bounds correspond to the situations where $\rho \gg \rho_{\mathrm{f}}$ (negligible hydrodynamic backflow) and $\rho \ll \rho_{\mathrm{f}}$ (maximum hydrodynamic backflow), respectively.

\subsection{Newtonian liquid}\label{subsect:Newton}

{\footnotesize{

\begin{table}
\caption{\label{tab:parameters}Values of the density ratio $\rho / \rho_{\mathrm{f}}$ used in the numerical simulation of Eq. (\ref{eq:GLENewton}) and corresponding values of the dimensionless parameters $\tau_{\mathrm{f}} / \tau_c$ and $\sqrt{\tau_c} a_{\pm}$ representing examples of some possible experimental systems at common room temperature.}
\footnotesize
\begin{tabular}{c|c|c|c|l}
\br
$\rho/\rho_{\mathrm{f}}$ & $\tau_{\mathrm{f}}/\tau_c$ & $\sqrt{\tau_c} a_{+}$ & $\sqrt{\tau_c} a_{-}$ & Examples \\
\mr
22.5&  0.1957 &  $0.2212 + 0.9752i$ &  $0.2212 - 0.9752i$ & Gold particle ($\rho = 19300 \, \mathrm{kg} \,\mathrm{m}^{-3}$) \\ 
&&&&in ethanol/water at 75\%~wt ($\rho_{\mathrm{f}} = 855 \, \mathrm{kg} \,\mathrm{m}^{-3}$) \\
2.5  &1.5&$0.6124 + 0.7906i$&$0.6124 - 0.7906i$ &Silica particle ($\rho = 2500 \, \mathrm{kg} \,\mathrm{m}^{-3}$) \\
&&&&in pure water ($\rho_{\mathrm{f}} = 1000 \, \mathrm{kg} \,\mathrm{m}^{-3}$)  \\
0.6&4.0909 &$1.1621$&$0.8605$&Polyethylene particle ($\rho =  850 \, \mathrm{kg} \,\mathrm{m}^{-3}$) \\
&&&&in glucose syrup ($\rho_{\mathrm{f}} =  1430 \, \mathrm{kg} \,\mathrm{m}^{-3}$)  \\
\br
\end{tabular}
\end{table}
}}

\normalsize

\begin{figure}
    \centering
\includegraphics[width=0.95\columnwidth]{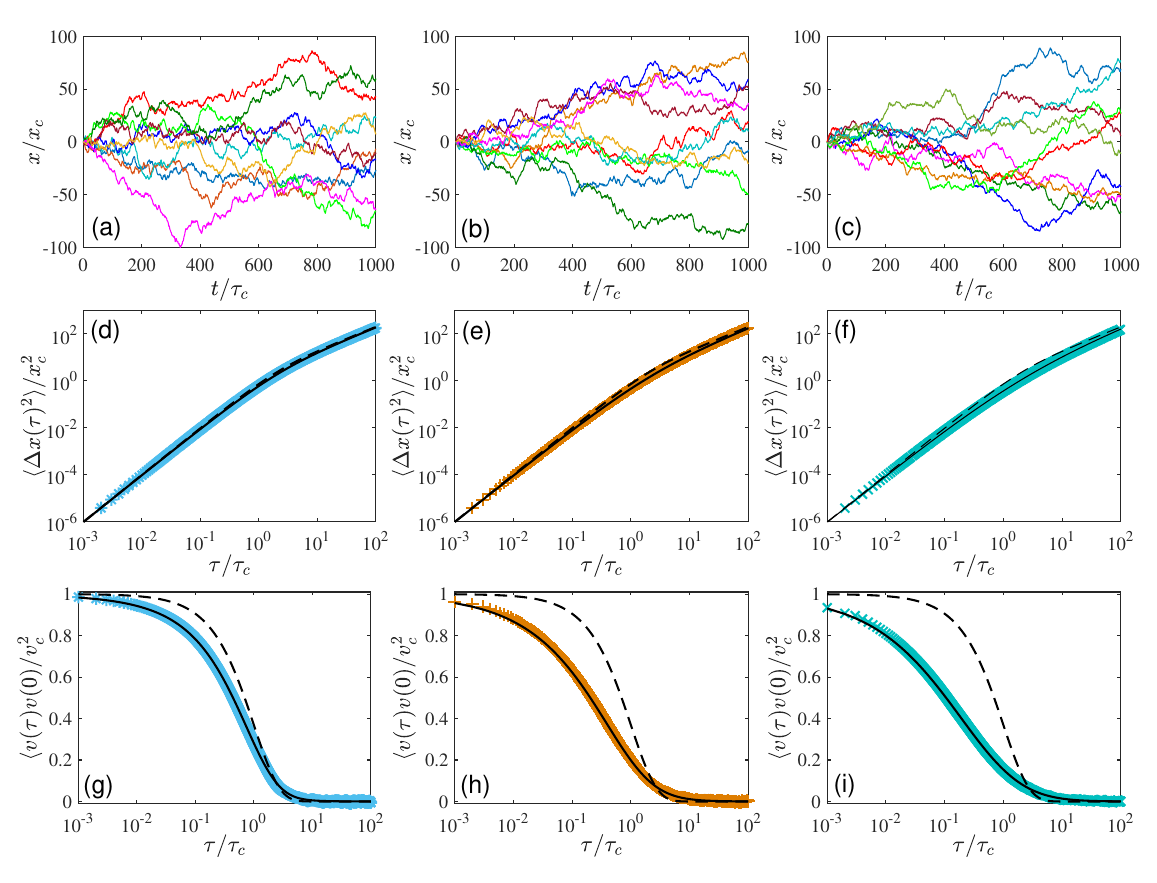}
\caption{Examples of the stochastic time evolution of the one-dimensional position of a spherical particle in an incompressible viscous fluid obtained by numerical simulation of Eq. (\ref{eq:GLENewton}) with initial conditions $x(0) = 0$ and $v(0) = v_0$, where $v_0$ is drawn from a normal distribution with mean 0 and variance $k_B T/M$, for some representative values of the ratio of the particle's to the fluid's mass density; (a) $\rho / \rho_{\mathrm{f}} = 22.5$, (b) $\rho / \rho_{\mathrm{f}} = 2.5$, and (c) $\rho / \rho_{\mathrm{f}} =0.6$. Different colors depict distinct initial conditions. The solid lines represent Eqs.~(\ref{eq:MSDNewton}) and (\ref{eq:psiNewton}) that include the effect of the Basset-Boussinesq force on the particle dynamics, whereas the dashed lines correspond to the case without hydrodynamic backflow.}\label{fig:1}
\end{figure}

First, we analyze the statistics of the fluctuations of the energy exchanged as heat between a spherical particle and a Newtonian fluid at constant temperature $T$, which is characterized by a constant viscosity $\eta$ that is frequency-independent, i.e. $\hat{\eta}(\omega) = \eta_{\infty} = \eta$, whose corresponding relaxation modulus is $G(t) = 2\eta \delta(t)$ for $t \ge 0$ and $G(t) = 0$ for $t < 0$. In such a case where elasticity is absent in the liquid, the Fourier transform of the memory kernel defined in Eq. (\ref{eq:GSFour}) takes the form 
\begin{equation}\label{eq:GammaNewton}
	\hat{\Gamma}(\omega) = 6\pi a \eta + 6\pi a^2 \sqrt{i\omega \rho_{\mathrm{f}} \eta}, 
\end{equation}
whose inverse Fourier transform yields the following expression for the generalized Langevin equation (\ref{eq:GLE}) describing the particle motion
\begin{equation}\label{eq:GLENewton}
    \dot{v}(t) = -\frac{1}{\tau_c} v(t) -\frac{1}{\tau_c} \sqrt{\frac{\tau_{\mathrm{f}}}{\pi}}  \int_0^t  dt' \, \frac{\dot{v}(t') } {\sqrt{t - t'}} + \frac{1}{M}\zeta(t).
\end{equation}
As described by Eqs. (\ref{eq:Lapsolvel}), (\ref{eq:Lappsi}), and (\ref{eq:VACF}), the corresponding stationary VACF, $\langle v(\tau) v(0) \rangle_{eq} = v_c^2 \psi(\tau)$, can be found by computing the Laplace transform of each term in Eq.~(\ref{eq:GLENewton}), which yields~\footnote{Note that because the memory kernel $\Gamma(t-t')$ is causal, its Fourier transform given in Eq. (\ref{eq:GammaNewton}) is simply related to its Laplace transform in Eq. (\ref{eq:Lappsi}) by $\hat{\Gamma}(\omega) = \tilde{\Gamma}(s = i\omega)$, thus directly leading to Eq. (\ref{eq:LappsiNewton}).}
\begin{equation}\label{eq:LappsiNewton}
	\tilde{\psi}(s) = \frac{1}{s + \frac{\sqrt{\tau_{\mathrm{f}}}}{\tau_c} \sqrt{s} +  \frac{1}{\tau_c}},
\end{equation}
whose inverse Laplace transform is known to possess the following analytic formula \cite{Hinch1975,Clercx1992,Mainardi1996}
\begin{equation}\label{eq:psiNewton}
   \psi(\tau)  = 
    \frac{1}{a_+ - a_-}\left[a_+ e^{a_+^2 \tau} \mathrm{erfc} \left( a_+ \sqrt{\tau}\right) - a_- e^{a_-^2 \tau} \mathrm{erfc} \left( a_- \sqrt{\tau}\right)\right].
\end{equation}
In Eq. (\ref{eq:psiNewton}), the constant parameters $a_{+}$ and $a_-$ are explicity given by
\begin{equation}\label{eq:apm}
	a_{\pm} = \frac{1}{2 \sqrt{\tau_{\mathrm{f}}}} \frac{1 \pm \sqrt{1 - 4 \frac{\tau_c}{\tau_{\mathrm{f}}}} }{\frac{\tau_c}{\tau_{\mathrm{f}}}},
\end{equation}
where $\mathrm{erfc}(z) = \frac{2}{\sqrt{\pi}} \int_z^{\infty} du \, e^{-u^2}$ stands for the complementary error function. Note that, for $\tau_{\mathrm{f}} \ge 4 \tau_c$, which corresponds to $\rho / \rho_{\mathrm{f}} \le 5/8$ according to Eq.~(\ref{eq:tauf}), the 
parameters $a_{\pm}$ defined in Eq. (\ref{eq:apm}) are real-valued, whereas they are complex for $\tau_{\mathrm{f}} < 4 \tau_c$, i.e. $\rho / \rho_{\mathrm{f}} > 5/8$. As shown in \cite{Makris2021}, the behavior of the VACF as a function of $\tau$ described by Eq. (\ref{eq:psiNewton}), which is real-valued, is continuous for all $0 \le \rho / \rho_{\mathrm{f}} < \infty$ even if the imaginary parts of the parameters $a_{\pm}$ change from zero to non-zero with increasing $\rho / \rho_{\mathrm{f}}$ when crossing the specific value $\rho / \rho_{\mathrm{f}} = 5/8$. Moreover, the asymptotic behavior of the VACF given by Eq. (\ref{eq:psiNewton}) for large $\tau$ exhibits a power-law tail with exponent~$-3/2$  
\begin{equation}\label{eq:VACFlong}
	\langle v(\tau) v(0) \rangle_{eq} \approx  \frac{v_c^2 \tau_c}{2\sqrt{\pi} \tau_{\mathrm{f}}} \left( \frac{\tau}{\tau_{\mathrm{f}}} \right)^{-3/2},
\end{equation}
which reflects the long-ranged hydrodynamic memory induced by the Basset-Boussinesq force ~\cite{Widom1971,Hauge1973}.
Furthermore, from the  VACF expressed by means of the function $\psi(\tau)$ in Eq. (\ref{eq:psiNewton}), an analytic expression for the mean-squared displacement of the particle position can be derived in a straightforward fashion \cite{Clercx1992,Weitz1989}
{\footnotesize{
\begin{eqnarray}\label{eq:MSDNewton}
  \fl	\langle \Delta x(\tau)^2 \rangle &  =  & 2 v_c^2 \int_0^{\tau} dt \int_0^{t} dt' \psi(t'), \nonumber\\
	\fl	& = &2x_c^2 \left\{  \frac{\tau_{\mathrm{f}}}{\tau_c} - 1 + \frac{\tau}{\tau_c} - 2 \sqrt{\frac{\tau_{\mathrm{f}}\tau}{\pi \tau_c^2}  }  + \frac{1}{\tau_{c}^2(a_+ - a_-)}\left[ \frac{e^{ a_+^2 \tau } {\mathrm{erfc}} \left( a_+ \sqrt{\tau} \right)}{a_+^3} - \frac{e^{ a_-^2 \tau } {\mathrm{erfc}} \left( a_- \sqrt{\tau} \right)}{a_-^3} \right] \right\},
\end{eqnarray}}}
which exhibits the asymptotic behavior $\langle \Delta x(\tau)^2 \rangle \approx 2 v_c^2 \tau_c \left( \tau - {2}
 \sqrt{\tau_{\mathrm{f}} \tau/\pi} \right)$ for sufficiently large $\tau$, thus revealing again the retarding effect of the  hydrodynamic backflow on the particle dynamics to reach the diffusive behavior $\langle \Delta x(\tau)^2 \rangle \approx 2 v_c^2 \tau_c \tau$ with diffusion coefficient $v_c^2 \tau_c = \frac{k_B T}{6\pi a \eta}$.

\begin{figure}
    \centering
\includegraphics[width=0.975\columnwidth]{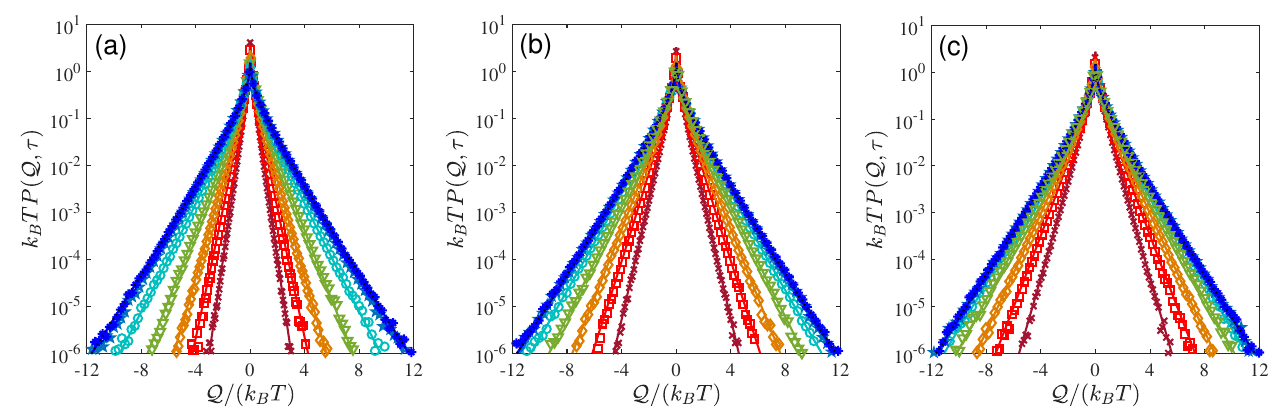}
\caption{Probability density function of the energy exchanged between a freely moving spherical particle and the surrounding viscous liquid for some representative values of the ratio of the particle's to the fluid's mass density: (a) $\rho / \rho_{\mathrm{f}} = 22.5$, (b) $\rho / \rho_{\mathrm{f}} = 2.5$, and (c) $\rho / \rho_{\mathrm{f}} =0.6$. The symbols represent probability densities that are computed from the numerically simulated trajectories during time intervals of distinct duration: $\tau = 0.003 \tau_c$ ($\times$), $\tau = 0.01 \tau_c$ (\scalebox{0.6}{$\square$}), $\tau = 0.03 \tau_c$ ($\diamond$), $\tau = 0.1 \tau_c$ ($\triangledown$), $\tau = 0.3 \tau_c$ ($\circ$), $\tau =  \tau_c$ ($\star$), and $\tau = 3 \tau_c$ ($*$). The solid lines represent the analytic expression given in Eq.~(\ref{eq:pdfQ}) for the corresponding values of $\tau$, whereas the dotted line depicts the asymptotic function for $\tau \rightarrow \infty$ in Eq.~(\ref{eq:limitpdfQ}).}\label{fig:2}
\end{figure}

To verify the main results of Section \ref{sect:thermo} on the statistics of the energy exchanged in a Newtonian liquid, we numerically simulate independent trajectories of the particle position, $x(t)$, evolving stochastically in time $t \ge 0$ according to Eq. (\ref{eq:GLENewton}) with initial condition $x(0) = 0, v(0) = v_0$, where $v_0$ is a stochastic variable with the Boltzmann distribution shown in Eq.~(\ref{eq:BD}) that describes the thermal equilibrium state of the system at $t = 0$. To this end, we implement a Markovian embedding method \cite{Siegle2010,Siegle2011}, which consists of an approximation of the non-Markovian part of the memory kernel described in Eq. (\ref{eq:GammaNewton}), i.e. the second term related to the Basset-Boussinesq force, as a finite sum of exponentially decaying functions. Thus, the non-Markovian dynamics described by Eq. (\ref{eq:GLENewton}) is simply a projection of a multidimensional Markovian dynamics of dimension $d = n+2$, where $n$ is the number of auxiliary variables associated to each exponential in the approximation. This method has the adventage to generate the thermal noise $\zeta(t)$ as a weighted finite sum of white noises satisfying the equilibrium properties described by Eq.~(\ref{eq:FD2nd}). More details on the numerical simulation of Eq. (\ref{eq:GLENewton}) using $n = 13$, where all the relevant physical quantities are rescaled by the characteristic scales defined in Eqs. (\ref{eq:tauc})-(\ref{eq:Fc}), are provided in Appendix \ref{app:simul}. The simulations are carried out for three different values of the density ratio $\rho / \rho_{\mathrm{f}}$ that represent possible experimental conditions, whose corresponding values of the parameters $\tau_{\mathrm{f}} / \tau_c$ and $\sqrt{\tau_c} a_{\pm}$ given by Eqs. (\ref{eq:tauf}) and (\ref{eq:apm}), respectively, are listed in Table \ref{tab:parameters}. Note that, while the case $\rho / \rho_{\mathrm{f}} = 2.5$ represents a commonly encountered situation in typical soft matter experiments, the values  $\rho / \rho_{\mathrm{f}} = 22.5$ and $\rho / \rho_{\mathrm{f}} = 0.6$ are particular examples of physical systems where the contrast between the mass density of the particle and that of the surrounding fluid are extreme. For each value of $\rho/\rho_{\mathrm{f}}$, 100 independent trajectories starting from the abovementioned initial conditions are simulated with a time-step $\Delta t = 10^{-4} \tau_c$, a total duration of $10^3 \tau_c$ and then sampled at a frequency $10^3 \tau_c^{-1}$, thus amounting to $10^8$ data points over which all the average quantities characterizing the dynamics of the system are computed. Some examples of these stochastic trajectories are shown in Fig. \ref{fig:1}(a)-(c), where different colors represent independent initial conditions. As shown in Figs.~\ref{fig:1}(d)-(f), such trajectories exhibit mean-squared displacements quantitatively described by Eq. (\ref{eq:MSDNewton}), which are depicted as solid lines, and deviate from the well known expression $2v_c^2 \tau_c \left[ \tau - \tau_c \left( 1 - e^{-t /\tau_c} \right)\right]$ for the mean-squared displacement of a particle freely moving in a fluid in absence of hydrodynamic backflow (dashed lines). This effect becomes more evident in the behavior of the VACF with decreasing values of $\rho / \rho_{\mathrm{f}}$, where significant deviations from the exponential decay $v_c^2 e^{-\tau/\tau_c}$ occurring without hydrodynamic memory (dashed lines) can be observed in Figs.~\ref{fig:1}(g)-(i) over the full time interval domain $10^{-3} \tau_c \le \tau \le 10^2 \tau_c$, whereas an excellent agreement is found with the expression given in Eq. (\ref{eq:psiNewton}) including the action of hydrodynamic backflow (solid lines).

\begin{figure}
    \centering
\includegraphics[width=0.95\columnwidth]{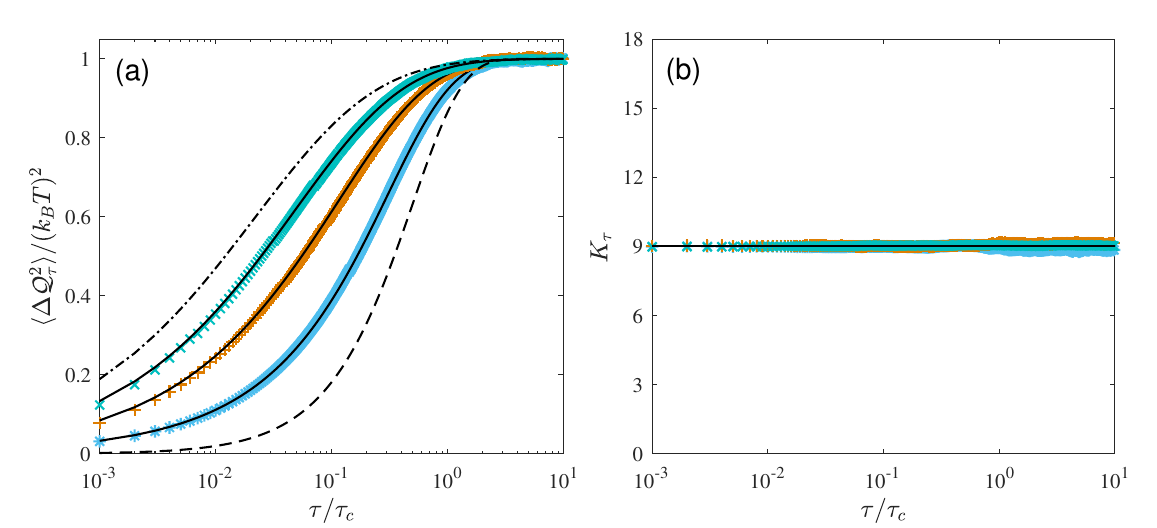}
\caption{(a) Dependence of the variance of the energy transferred between a freely moving spherical particle and the surrounding viscous liquid, $\langle \Delta \mathcal{Q}_{\tau}^2\rangle$, on the duration of the time interval over which the energy exchange takes place,  $\tau$, for some representative values of the mass-density ratio: $\rho / \rho_{\mathrm{f}} = 22.5$ ($*$), $\rho / \rho_{\mathrm{f}} = 2.5$ ($+$), and $\rho / \rho_{\mathrm{f}} =0.6$ ($\times$). The solid lines represent the general expression of Eq. (\ref{eq:varianceq}) for $\langle \Delta \mathcal{Q}_{\tau}^2\rangle$, which are computed using the function $\psi(\tau)$ given in Eq. (\ref{eq:psiNewton}). The dashed line corresponds to the behavior of the variance of the exchanged energy without including the effect of hydrodynamic backflow ($\tau_{\mathrm{f}} = 0$), whereas the dotted-dashed line depicts the upper bound of maximum hydrodynamic memory  ($\tau_{\mathrm{f}} = 9 \tau_c$) shown in (\ref{eq:intervalvariance}). (b) Dependence of the kurtosis of the energy transferred between a freely moving spherical particle and the surrounding viscous liquid on the duration of the time interval $\tau$, for $\rho / \rho_{\mathrm{f}} = 22.5$ ($*$), $\rho / \rho_{\mathrm{f}} = 2.5$ ($+$), and $\rho / \rho_{\mathrm{f}} =0.6$ ($\times$). The solid line represents the constant value $K_{\tau} = 9$ theoretically predicted by Eq.~(\ref{eq:kurt}).}\label{fig:3}
\end{figure}

Using the simulated trajectories including the effect of the Basset-Boussinesq force for the three distinct values of the density ratio $\rho / \rho_{\mathrm{f}}$,  by means of Eq. (\ref{eq:heat}) we compute the fluctuations of the energy exchanged during a time interval $\tau$, $\mathcal{Q}_{\tau}$, as well as their normalized histograms, which correspond to their probability density functions. For each value of $\rho/ \rho_{\mathrm{f}}$, in Figs. \ref{fig:2}(a)-(c) we plot such probability densities for different time intervals that cover four orders of magnitude from $\tau = 0.003 \tau_c$ to $\tau = 3 \tau_c$, which are depicted by distinct symbols and colors. We verify that for all the analyzed values of $\tau$ and within the range $|\mathcal{Q}| \le 12k_B T$ extending over six orders of magnitude of values of the probability density function of $\mathcal{Q}_{\tau}$, namely $10^{-6} (k_B T)^{-1} \le P(\mathcal{Q},\tau) < 10^1 (k_B T)^{-1}$, the analytic expression of $P(\mathcal{Q},\tau)$ given in Eq.~(\ref{eq:pdfQ}) in terms of the zeroth order modified Bessel function of the second kind describes accurately the statistics of the energy transfer of the simulated stochastic dynamics. To perfom such a comparison, for each value of $\tau$ we calculated the corresponding numerical value of the function $\psi(\tau)$ using the expression presented in Eq. (\ref{eq:psiNewton}), which was then substituted into the theoretically derived formula of $P(\mathcal{Q},\tau)$ given in Eq.~(\ref{eq:pdfQ}) to finally obtain the solid curves plotted in Fig.~\ref{fig:2}, where same colors represent the same values of $\tau$. We also find that, for the three values of the density ratio $\rho / \rho_{\mathrm{f}}$, the shape of $P(\mathcal{Q},\tau)$ becomes independent of $\tau$ for sufficiently large $\tau$ and converges to the function $P(\mathcal{Q},\tau \rightarrow \infty)$ given in Eq. (\ref{eq:limitpdfQ}), as can be observed in Figs. \ref{fig:2}(a)-(c) for $\tau \gtrsim \tau_c$ , at which $P(\mathcal{Q},\tau)$ is practically indistinguishable from $P(\mathcal{Q},\tau \rightarrow \infty)$. Moreover, by comparing the width of the probability densities traced in Figs. \ref{fig:2}(a), (b) and (c), we can realize that the strength of the Basset-Boussinesq force significantly impacts the amplitude of the fluctuations of the exchanged energy during short and intermediate time intervals relative to momentum relaxation time of the particle, i.e. $0 < \tau \lesssim \tau_c$, whose overall effect is to broaden their corresponding probability density  with decreasing $\rho / \rho_{\mathrm{f}}$ for a given value of $\tau$. Indeed, this is verified in Fig. \ref{fig:3}(a), where we plot as symbols the dependence  on the time span $\tau$ of the variance of $\mathcal{Q}_{\tau}$, $\langle \Delta \mathcal{Q}_{\tau}^2\rangle$, which is computed for the three sets of trajectories simulated for each value of $\rho / \rho_{\mathrm{f}}$. We observe that the smaller the value of $\rho / \rho_{\mathrm{f}}$, i.e. the stronger the Basset-Boussinesq force quantified by the ratio $\tau_{\mathrm{f}} / \tau_c$ whose values are listed in Table \ref{tab:parameters}, the larger the corresponding variance $\langle \Delta \mathcal{Q}_{\tau}^2\rangle$ with respect to the curve $(k_B T)^2 \left( 1 - e^{-2\tau / \tau_c} \right)$  over $0 < \tau \lesssim \tau_c$, which is drawn as a dashed line in Fig. \ref{fig:3}(a) and corresponds to the case without hydrodynamic backflow. This results provides evidence that the hydrodynamic memory induced by the Basset-Boussinesq force enhances the energy transfer process between the particle and the surrounding Newtonian liquid, both the energy absorption from the bath and the release into it, on time-scales that are shorter than the momentum relaxation time of the particle. Furthermore, in Fig. \ref{fig:3}(a) we also check that the theoretically derived expression in Eq. (\ref{eq:varianceq}) for the variance $\langle \Delta \mathcal{Q}_{\tau}^2\rangle$ in terms of the function $\psi(\tau)$, which represents the normalized VACF of the particle, quantitatively agrees with the numerical results obtained from the simulation of the stochastic dynamics following Eq. (\ref{eq:GLENewton}), as shown by the solid lines Fig. \ref{fig:3}(a). It should be noted that for an arbitrary value of the density ratio $\rho / \rho_{\mathrm{f}}$, the curve representing the dependence of the variance $ \langle \Delta \mathcal{Q}_{\tau}^2\rangle$ on $\tau$ is bounded on $0 < \tau \lesssim \tau_c$ by
{\footnotesize{
\begin{eqnarray}\label{eq:intervalvariance}
	\fl	 1 - e^{-\frac{2\tau }{ \tau_c}} & < &\frac{ \langle \Delta \mathcal{Q}_{\tau}^2\rangle } {(k_B T)^2} \nonumber\\
	\fl & < & 1 - \frac{e^{\frac{7\tau}{\tau_c}} \left[ \left(3 + \sqrt{5} \right)e^{ \frac{3\sqrt{5}}{2} \frac{\tau}{\tau_c}} \mathrm{erfc} \left( \frac{3 + \sqrt{5}}{2} \sqrt{\frac{\tau}{\tau_c}} \right) -  \left(3 - \sqrt{5} \right)e^{- \frac{3\sqrt{5}}{2} \frac{\tau}{\tau_c}} \mathrm{erfc} \left( \frac{3 - \sqrt{5}}{2} \sqrt{\frac{\tau}{\tau_c}} \right)\right]^2}{20}.
\end{eqnarray}}}
The lower and upper bounds in the inequiality (\ref{eq:intervalvariance}) correspond to the cases $ \tau_{\mathrm{f}} = 0$ (no hydrodynamic memory) and $\tau_{\mathrm{f}} = 9 \tau_c$ (maximum hydrodynamic memory when $\rho / \rho_{\mathrm{f}} \rightarrow 0$), where the behavior of  $\langle \Delta \mathcal{Q}_{\tau}^2\rangle$ for most of the physical situations that can be achieved in experiments lie between the curves corresponding to $ \tau_{\mathrm{f}} \approx 0.2 \tau_c$ and $\tau_{\mathrm{f}} \approx 4 \tau_c$. It should also be pointed out that, as described by Eq. (\ref{eq:psiNewton}) and illustrated in Figs \ref{fig:1}(d)-(f), the VACF of the particle moving in a Newtonian liquid is a monotonically decreasing function of the time lag $\tau$. Therefore, as shown in Fig. \ref{fig:3}(a), the variance of $\mathcal{Q}_{\tau}$ increases monotonically with increasing $\tau$, then eventually saturating for sufficiently large $\tau$ to the value $\langle \Delta \mathcal{Q}_{\tau \rightarrow \infty}^2\rangle = (k_B T)^2$, whose square root represents the maximum value of the characteristic amount of energy that can be exchange with the thermal bath. Note that despite the slow approach to this limit as a power law with exponent $-3$ according to the asymptotic behavior for $\tau \gg \tau_c$ of the function given in Eq. (\ref{eq:VACFlong}), i.e. $(k_B T)^2 - \langle \Delta \mathcal{Q}_{\tau}^2\rangle \approx (k_B T)^2 \frac{\tau_{\mathrm{f}}}{4\pi \tau_c} \left( \frac{\tau}{\tau_c}\right)^{-3}$, even in the extreme case of the strongest hydrodynamic memory analyzed here ($\tau_{\mathrm{f}} \approx 4.0909 \tau_c$), $\langle \Delta \mathcal{Q}_{\tau}^2\rangle$ reaches 97.7\% of the value $(k_B T)^2$ at $\tau = \tau_c$, whereas at $\tau = 5 \tau_c$ it amounts to 99.9\%. Consequently, in most cases of actual physical systems, it can be considered that the variance of the exchanged energy has virtually attained its maximum value  for $\tau \gtrsim \tau_c$.
To end this Subsection, in Fig.~\ref{fig:3}(b) we show that the kustosis of $\mathcal{Q}_{\tau}$, depicted as symbols, is independent of $\tau$ and equal to the constant value $K_{\tau} = 9$ that is theoretically predicted by Eq. (\ref{eq:kurt}) for the three distinct values of the density ratio. This result attests that for all the analyzed values of $\tau$, the shape of the numerically determined probability density function of  $\mathcal{Q}_{\tau}$ is accurately described by the zeroth order modified Bessel function of the second kind given in Eq.~(\ref{eq:pdfQ}).

\subsection{Viscoelastic liquid}\label{subsect:VE}

\begin{figure}
    \centering
\includegraphics[width=0.975\columnwidth]{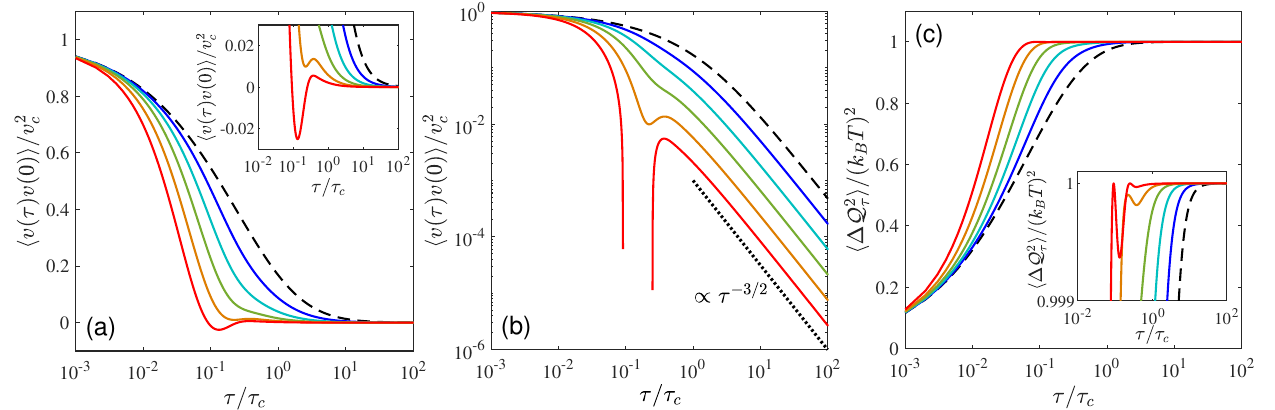}
\caption{(a) Semi-log plot of the normalized VACF of a spherical particle moving in a viscoelastic liquid of the same mass density ($\tau_{\mathrm{f}} = 3 \tau_c$), whose stress relaxation modulus is given by Eq. (\ref{eq:modulusMaxwell}) with relaxation time $\lambda = 0.1 \tau_c$ and different values of the viscosity ratio: $\eta_0 / \eta_{\infty} = 2$ (blue solid line),  $\eta_0 / \eta_{\infty} = 4$ (turquoise solid line),  $\eta_0 / \eta_{\infty} = 8$ (green solid line),  $\eta_0 / \eta_{\infty} = 16$ (orange solid line), and  $\eta_0 / \eta_{\infty} = 32$ (red solid line). The black dashed line represents the normalized VACF given in Eq.~(\ref{eq:psiNewton}) in the case of a purely viscous liquid, which corresponds to $\eta_0 / \eta_{\infty} =  1$. Inset: expanded view of the main plot around $\langle v(\tau) v(0) \rangle_{eq} = 0$. (b) Log-log representation of the normalized VACF curves plotted in Fig. \ref{fig:4}(a). Same color and linestyle codes as in Fig. \ref{fig:4}(a). The dotted line represents the power-law decay $\langle v(\tau) v(0) \rangle_{eq} \propto \tau^{-3/2}$ of the VACF resulting from hydrodynamic backflow. (c) Dependence on $\tau$ of the variance of the energy exchanged between the particle and the viscoelastic liquid for $\tau_{\mathrm{f}} = 3 \tau_c$, $\lambda = 0.1 \tau_c$, and different values of $\eta_0 / \eta_{\infty}$. Same color and linestyle codes as in Fig. \ref{fig:4}(a). Inset: expanded view of the main plot around $\langle \Delta \mathcal{Q}_{\tau}^2 \rangle / (k_B T)^2 = 1$.}\label{fig:4}
\end{figure}

Since we have verified in the previous Subsection that the statistical properties of the heat exchanged between a Brownian particle and a surrounding Newtonian liquid is quantitatively described by the general equations derived in Section \ref{sect:thermo} involving the VACF of the particle, we can safely employ them to investigate the effect of the viscoelasticity of the liquid in addition to the Basset-Boussinesq force on the energy transfer process. For the sake of simplicity and without loss of generality, we consider the following linear viscoelastic model for the stress relaxation modulus of the liquid
\begin{equation}\label{eq:modulusMaxwell}
   G(t) = \left\{
    \begin{array}{ll}
   2\eta_{\infty}\delta(t) + \frac{\eta_0 - \eta_{\infty}}{\lambda} \exp \left( -\frac{t}{\lambda} \right), & \,\,\,\,\,  t \ge 0 ,\\
   0, & \,\,\,\,\,  t < 0.
    \end{array} \right.
\end{equation}
In Eq. (\ref{eq:modulusMaxwell}), apart from the intantaneous response of the solvent of viscosity $\eta_{\infty}$ described by the term proportional to the Dirac delta function, the expression also includes an exponential Maxwell-like term that accounts for the slower stress relaxation of the embedded macromolecular microstructure of the liquid, which endows it with an extra viscosity $\eta_0 - \eta_{\infty} > 0$ and a non-zero relaxation time $\lambda > 0$. Therefore, for this model the zero-shear viscosity of the liquid and its elastic shear modulus are $\eta_0 = \int_0^{\infty} dt\, G(t) > \eta_{\infty}$ and $(\eta_0 - \eta_{\infty})/\lambda > 0$, respectively \cite{bird1987dynamics}. The Laplace transform of $G(t)$ is 
\begin{equation}\label{eq:etaMaxwell}
	\tilde{\eta}(s) \equiv \tilde{G}(s) = \eta_{\infty} + \frac{\eta_0 - \eta_{\infty}}{1 + \lambda s},
\end{equation}
whereas its Fourier transform defines a complex viscosity at the real-valued frequency $\omega$ that can be simply determined by taking $s = i\omega$ in Eq.~(\ref{eq:etaMaxwell}), i. e. $\hat{\eta}(\omega) = \tilde{\eta}( s = i\omega) = \eta_{\infty} + \left( \eta_0 - \eta_{\infty}\right)/\left( 1 + i\omega \lambda \right)$ because $G(t)$ is causal. Note that, besides the characteristic scales defined in Eqs. (\ref{eq:tauc})-(\ref{eq:tauf}), in this case the dimensionless parameters $\eta_0 / \eta_{\infty}$ and $\lambda / \tau_c$ are also needed to describe the energy exchange process between the particle and the viscoelastic liquid, where the former weighs the additional dissipation due to the macromolecular microstructure of the fluid, whereas the latter quantifies the non-Markovianity due to its non-instantaneous stress relaxation.

The causality of the memory kernel defined by its Fourier transform in Eq. (\ref{eq:GSFour}) allows us to express its Laplace transform as $\tilde{\Gamma}(s) =  6\pi a \tilde{\eta}(s) + 6\pi a^2 \sqrt{s \rho_{\mathrm{f}} \tilde{\eta}(s)}$, which, by use of Eqs. (\ref{eq:Lappsi}) and (\ref{eq:etaMaxwell}), leads to the following expression for the Laplace transform of the function $\psi(\tau)$ that appears in the general formula of Eq. (\ref{eq:pdfQ}) for the probability density of the exchanged energy
\begin{equation}\label{eq:LappsiMaxwell}
	\tilde{\psi}(s) = \frac{\tau_c}{\tau_c s + 1 + \frac{{\eta_0}/{\eta_{\infty}}-1}{1 + \lambda s} + \sqrt{\tau_{\mathrm{f}} s}\sqrt{1 + \frac{{\eta_0}/{\eta_{\infty}} -1}{1 + \lambda s}}},
\end{equation}
whose inverse Laplace transform cannot be analytically expressed in general apart from some specific limiting cases presented later. Therefore, in order to study the properties of the fluctuations of the energy exchanged during a finite time interval $\tau$ through the behavior of the function $\psi(\tau)$, Eq.~(\ref{eq:LappsiMaxwell}) is numerically inverted in the following by means of the Talbot method \cite{Talbot1979}. In addition, for the sake of simplicity, we will focus on the situation where the bead and the fluid have the same mass density, i.e. $\tau_{\mathrm{f}} = 3 \tau_c$ , which is typical in experiments.

In Fig. \ref{fig:4}(a), we illustrate the behavior of the normalized VACF, $\langle v(\tau) v(0) \rangle_{eq}/v_c^2 = \psi(\tau)$ at a fixed value of the liquid's stress relaxation time ($\lambda/\tau_c = 0.1$) and varying the viscosity ratio $\eta_0/\eta_{\infty}$. As a reference, we start with the case $\eta_0 / \eta_{\infty} =1$ which is depicted as dashed lines in Fig. \ref{fig:4} and represents the situation where Eq. (\ref{eq:LappsiMaxwell}) reduces to Eq. (\ref{eq:LappsiNewton}), i.e. it coincides with the motion in a Newtonian liquid, in which case the normalized VACF is described by Eq. (\ref{eq:psiNewton}), thus exhibiting a monotonic decrease a function of $\tau$. Increasing the viscosity ratio gives rise to systematic deviations from the reference curve at $\eta_0 / \eta_{\infty} =1$, where the monotonic decay of the VACF becomes faster with increasing values of $\eta_0 / \eta_{\infty}$, as shown in Fig. \ref{fig:4}(a) for $\eta_0 / \eta_{\infty} =2$, $\eta_0 / \eta_{\infty} =4$, and $\eta_0 / \eta_{\infty} =8$. This can be attributed to the fact that the momentum relaxation time of the particle and the vortex diffusion time in the viscoelastic liquid are both effectively reduced with respect to the reference values given by Eqs. (\ref{eq:tauc}) and (\ref{eq:tauf}), respectively, because of an increase in the effective viscosity of the liquid with respect to the solvent viscosity $\eta_{\infty}$, which approaches the value $\eta_0$ at low frequencies according to Eq. (\ref{eq:etaMaxwell}), i.e at long time-scales. Consequently, for sufficiently large values of $\tau$, the decay of the VACF must asymptotically behave as the power law described by Eq. (\ref{eq:VACFlong}) with $\tau_c$ and $\tau_{\mathrm{f}}$ reduced by the amount $\eta_0/\eta_{\infty}$, which yields a decrease in the VACF by a factor equal to $\left(\eta_{\infty}/\eta_0\right)^{3/2}$. Indeed, this is verified in Fig. \ref{fig:4}(b), where the power-law behavior of the normalized VACF with the correct diminution can be clearly observed for the previoulsy mentioned values of $\eta_0 / \eta_{\infty}$. As a consequence, for these values of the viscosity ratio the variance of the exchanged energy, $\langle \Delta \mathcal{Q}_{\tau}^2\rangle$, is a monotonically increasing function of the time interval $\tau$, as can be seen in Fig. \ref{fig:4}(c), where the covergence to the limit value $\langle \Delta \mathcal{Q}_{\tau \rightarrow \infty}^2\rangle = (k_B T)^2$ occurs faster with larger $\eta_0/\eta_{\infty}$. This demonstrates that the effect of increasing the viscosity ratio, which physically corresponds to augment the concentration of macromolecules suspended in the solvent composing the liquid, is to boost the energy absorption by the particle and its dissipation into the surroundings in addition to the enhancement induced by Basset-Boussinesq force due to the solvent. Interestingly, we find that a further increase in the viscosity ratio gives rise to a non-monotonic behavior of the VACF as a function of $\tau$, as can be seen in Fig. \ref{fig:4}(a) for $\eta_0 / \eta_{\infty} = 16$ and $\eta_0 / \eta_{\infty} = 32$. Such a non-monotonic behavior of the VACF at sufficiently large $\eta_0 / \eta_{\infty}$ is manifested either as a local minimum at which the VACF is non-negative followed by a local maximum, e.g. for $\eta_0 / \eta_{\infty} = 16$, or as global minimum at which anticorrelations of the particle velocity happen, i.e. $\langle v(\tau) v(0) \rangle_{eq} < 0$, followed by a local maximum with positive velocity correlations, e.g. for $\eta_0 / \eta_{\infty} = 32$. The appearance of a non-monotonic behavior of the VACF, e.g. damped oscillations, has already been observed for generalized Langevin models that include a non-instantaneous memory kernel in the Stokes force but with the Basset-Boussinesq force only determined by a constant viscosity \cite{PhysRevE.88.040701,PhysRevE.89.012130,Vinales2020,BAKALIS2023128780}. However, it should be pointed out that here we include the same frequency dependence of the fluid viscosity given by Eq. (\ref{eq:etaMaxwell}) in both terms associated to the Stokes and the Basset-Boussinesq forces acting on the particle, in accordance with the correct hydrodynamic description of the system at all frequencies provided by Eq. (\ref{eq:GSFour}). It is important to notice that negative velocity correlations also take place in the case of a particle trapped in a purely viscous fluid by a harmonic potential \cite{Clercx1992,Huang2011,Kheifets2014}, where the role of the confining potential is to act as an elastic component of the system that transiently stores energy before fully transferring it to the particle or to the medium. Then, a qualitatively similar mechanism can be concieved for a particle freely moving in a viscoelastic liquid, where a sufficiently strong elasticity of the medium, quantified in this case by its elastic modulus $(\eta_0 - \eta_{\infty})/\lambda$, permits transient energy storage on intermediate time intervals, which must result in a non-monotonic behavior of the variance $\langle \Delta \mathcal{Q}_{\tau}^2\rangle$ before reaching the asymptotic value $(k_B T)^2$, In the inset of Fig \ref{fig:4}(c) we confirm that this scenario is actually observed for the variance of the energy exchanged between the particle and the viscoelastic liquid with $\eta_0 / \eta_{\infty} = 16$ and $\eta_0 / \eta_{\infty} = 32$, where a larger value of the viscosity ratio corresponding to a larger elastic modulus of the liquid at fixed $\lambda$, gives rise to a more pronounced dip in the shape of $\langle \Delta \mathcal{Q}_{\tau}^2\rangle$ as a function of $\tau$. This is consistent with time intervals over which the characteristic energy absorbed or released by the particle, quantified by the standard deviation $\sqrt{\langle \Delta \mathcal{Q}_{\tau}^2\rangle}$, diminishes with respect to the values at shorter durations. The decrease is due to the amount of energy that is stored in the elastic microstructure of the liquid surrounding the particle, which gradually relaxes with increasing values of $\tau$ to finally enable the particle to exchange the maximum amount of energy with the bath, whose characteristic value is $k_B T$. Note that, despite the development of such an intricate energetic behavior when the elasticity of the liquid is significant, for sufficiently large values of $\tau$ the power-law behavior of the VACF due the hydrodynamic memory is revealed as well as the rapid convergence of $\langle \Delta \mathcal{Q}_{\tau}^2\rangle$ to $(k_B T)^2$ for $\tau \gtrsim \tau_c$, as clearly shown in Figs. \ref{fig:4}(b) and (c), respectively, for all values of the viscosity ratio including $\eta_0 / \eta_{\infty} = 16$ and $\eta_0 / \eta_{\infty} = 32$.

\begin{figure}
    \centering
\includegraphics[width=0.975\columnwidth]{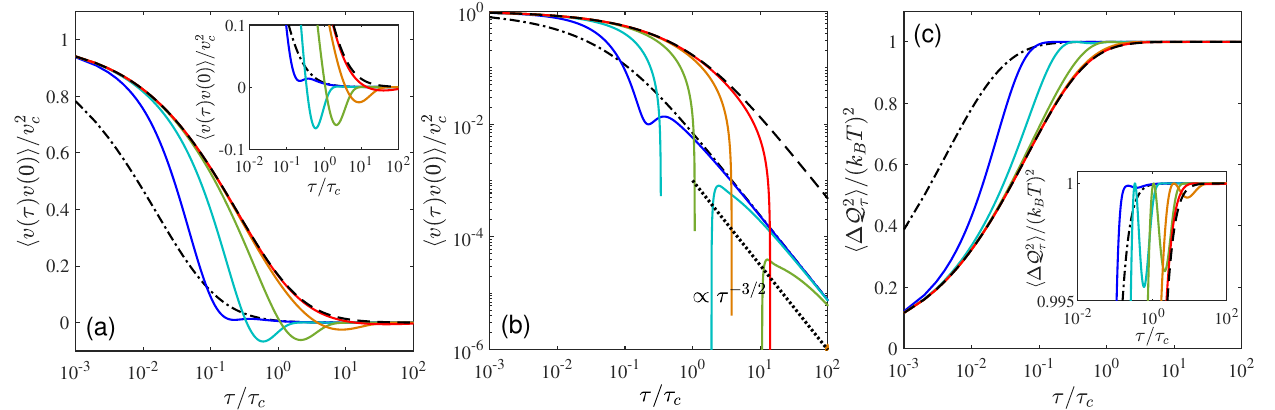}
\caption{(a) Semi-log plot of the VACF of a spherical particle moving in a viscoelastic liquid of the same mass density ($\tau_{\mathrm{f}} = 3 \tau_c$), whose stress relaxation modulus is given by Eq. (\ref{eq:modulusMaxwell}) with viscosity ratio $\eta_0 / \eta_{\infty} = 16$ and different values of the relaxation time: $\lambda = 0.1 \tau_c$ (blue solid line),  $\lambda = \tau_c$ (turquoise solid line),  $\lambda =10 \tau_c$ (green solid line),  $\lambda = 100 \tau_c$ (orange solid line), and  $\lambda = 1000 \tau_c$ (red solid line). The black dashed line represents the VACF given in Eq. (\ref{eq:psiNewton}) in the case of a Newtonian liquid of viscosity $\eta_{\infty}$, whereas the dotted-dashed line corresponds to the limit $\lambda \rightarrow 0$ at fixed $\eta_0/\eta_{\infty} = 16$. Inset: expanded view of the main plot around $\langle v(\tau) v(0) \rangle_{eq} = 0$. (b) Log-log representation of the VACF curves plotted in Fig. \ref{fig:5}(a). Same color and linestyle codes as in Fig. \ref{fig:5}(a). The dotted line represents the power-law decay $\langle v(\tau) v(0) \rangle_{eq} \propto \tau^{-3/2}$ of the VACF resulting from hydrodynamic backflow. (c) Dependence on $\tau$ of the variance of the energy exchanged between the particle and the viscoelastic liquid for $\tau_{\mathrm{f}} = 3 \tau_c$, $\eta_0 / \eta_{\infty} = 16$, and different values of $\lambda$. Same color and linestyle codes as in Fig. \ref{fig:5}(a). Inset: expanded view of the main plot around $\langle \Delta \mathcal{Q}_{\tau}^2 \rangle / (k_B T)^2 = 1$.}\label{fig:5}
\end{figure}

In Fig. \ref{fig:5}(a) we analyze the influence of the normalized stress relaxation time $\lambda / \tau_c$ on the behavior of the normalized VACF at a fixed viscosity ratio ($\eta_0 / \eta_{\infty} = 16$). To better understand the role of this parameter, we plot two reference curves, one corresponding to the limit $\lambda / \tau_c \rightarrow 0$, for which Eq. (\ref{eq:LappsiMaxwell}) reduces to the expression
\begin{equation}\label{eq:Lappsieta0}
	\tilde{\psi}(s) = \frac{1}{s + \frac{\eta_0}{\eta_{\infty} \tau_c} \sqrt{\frac{\eta_{\infty} \tau_{\mathrm{f}}}{\eta_0} s} + \frac{\eta_0}{\eta_{\infty}\tau_c}}
\end{equation}
whose inverse Laplace transform is depicted as a dotted-dashed line in Fig. \ref{fig:5}(a) and is analitically described by Eq. (\ref{eq:psiNewton}) with $\tau_c$ and $\tau_{\mathrm{f}}$ replaced by $\eta_{\infty}\tau_c / \eta_0$ and $\eta_{\infty} \tau_{\mathrm{f}} / \eta_0$, respectively, thus effectively representing the motion in a Newtonian fluid of viscosity $\eta_0 > \eta_{\infty}$. The second reference curve, which is traced as a dashed line in Fig. \ref{fig:5}(a), is associated to the limit of the VACF as $\lambda / \tau_c \rightarrow \infty$, whose analytic expression is simply given by Eq. (\ref{eq:psiNewton}), i.e. it can be regarded as the motion in the pure solvent with viscosity $\eta_{\infty}$. We find that for all the explored values of $\lambda / \tau_c$, even when the normalized VACF can be non-monotonic, its behavior at sufficiently short $\tau$ is very similar to that of a particle moving in plain solvent because the effective viscosity of the viscoelastic liquid approaches $\eta_{\infty}$ at high frequencies, i.e. at short time-scales. Nevertheless, deviations from this reference curve can be orderly detected depending on $\lambda / \tau_c$. For instance, for a very large stress relaxation time, e.g. $\lambda / \tau_c = 1000$, the behavior of the normalized VACF is very close to that of a particle in a Newtonian fluid of viscosity $\eta_{\infty}$, with negative velocity correlations of very small amplitude ocurring at rather large values of $\tau$. A further reduction of $\lambda / \tau_c$ has the effect to drop the value of the characteristic time at which the VACF decays first to zero, which is followed by a global minimum around which velocity anticorrelations appear, and then at larger $\tau$ the VACF exhibits a local maximum with positive velocity correlations that ultimately vanish as $\tau \rightarrow \infty$, as can be seen in Figs. \ref{fig:5}(a) and \ref{fig:5}(b) for  $\lambda / \tau_c = 100$,  $\lambda / \tau_c = 10$, and  $\lambda / \tau_c = 1$. Note that decreasing the value of $\lambda / \tau_c$ at constant $\eta_0 / \eta_{\infty} > 0$ signifies a larger elastic modulus of the liquid, $(\eta_0 - \eta_{\infty})/\lambda$, which results in a gradual increase of the amplitude of the anticorrelations of the particle velocity, as can be corroborated in Fig. \ref{fig:5}(a) for $\lambda / \tau_c = 1000$, $\lambda / \tau_c = 100$,  $\lambda / \tau_c = 10$, and  $\lambda / \tau_c = 1$. Therefore, for these values of $\lambda / \tau_c$ the transient energy storage by the liquid results in the non-monotonic behavior of the variance $\langle \Delta \mathcal{Q}_{\tau}^2 \rangle$ represented in Fig. \ref{fig:5}(c), where it can be observed that the larger the elastic modulus of the liquid, the larger the decline of the variance before growing again to reach the final value $(k_B T)^2$. Surprisingly, an additional decrease in $\lambda$ does not give rise to a further amplification of negative velocity correlations or a more pronounced drop of the variance of the exchanged energy at intermediate time intervals, but to the disappearance of negative values of the VACF and a significant decrease of the non-monotonic behavior of $\langle \Delta \mathcal{Q}_{\tau}^2 \rangle$, as shown in Figs. \ref{fig:5}(a) and \ref{fig:5}(c) for $\lambda / \tau_c = 0.1$. This effect can be explained by the fact that, if $\lambda$ is sufficiently short, then the friction felt by the particle must get closer to that set by $\eta_0$ so that elastic effects quickly fade out, thus leading a drastic decrease in the energy storage capability of the liquid and a fast convergence of  $\langle \Delta \mathcal{Q}_{\tau}^2 \rangle$ to the asymptotic value $(k_B T)^2$ for $\tau \gtrsim \tau_c$. Furthermore, in Fig. \ref{fig:5}(b) we show that, irrespective of the value of $\lambda > 0$, the power-law decay $\propto \tau^{-3/2}$ of the VACF takes place for sufficiently large values of $\tau$, whose amplitude is consistent with the motion in a liquid of viscosity $\eta_0$ (dotted-dashed line), as expected from the behavior of the viscosity of the liquid given in Eq. (\ref{eq:etaMaxwell}) at small frequencies, once elastic effects on the particle motion have vanished.

\begin{figure}
    \centering
\includegraphics[width=0.975\columnwidth]{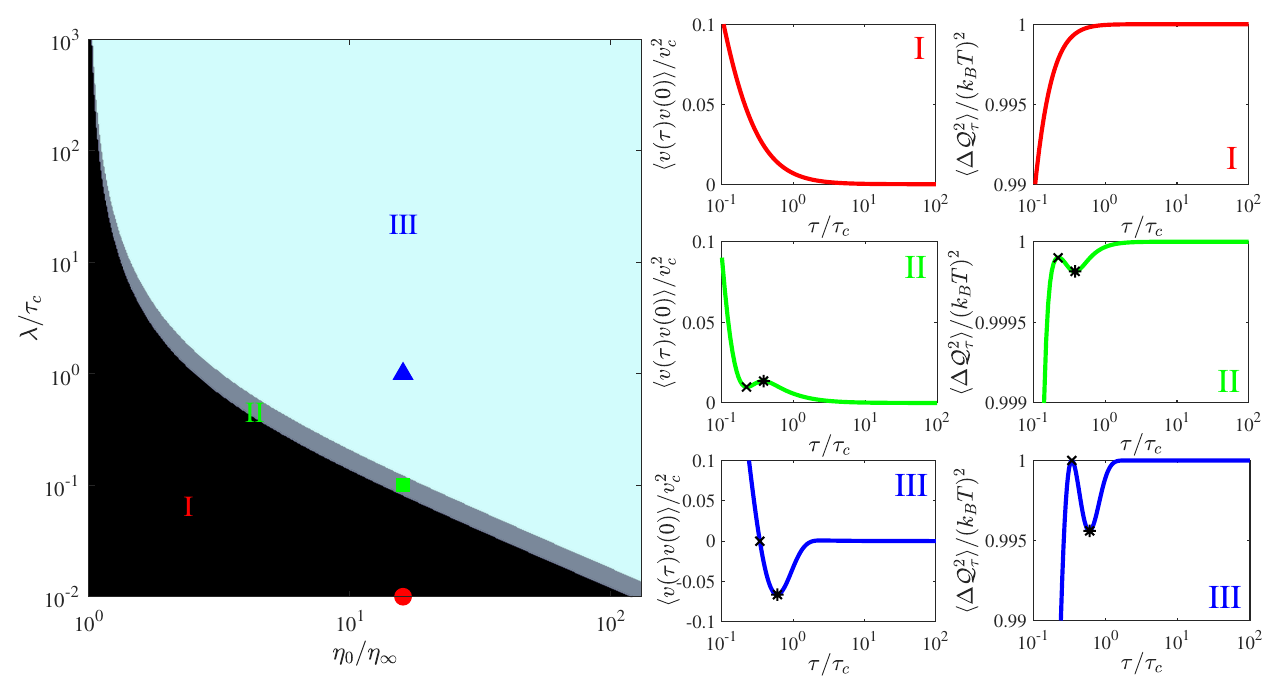}
\caption{Diagram of the three different types of behavior of the variance $\langle \Delta \mathcal{Q}_{\tau}^2 \rangle$ of the energy exchanged over a time interval of duration $\tau$ between a spherical particle and a surrounding Maxwell-type viscoelastic liquid of the same mass density ($\tau_{\mathrm{f}} = 3 \tau_c$), depending on the values of viscosity ratio $\eta_0/\eta_{\infty}$ and the normalized stress relaxation time $\lambda / \tau_c$: I) monotonic increase as a function of $\tau$ (black area), II) non-monotonic increase with a local maximum followed by a local minimum at $\tau_{\times}$ and $\tau_*$, respectively, such that $\langle \Delta \mathcal{Q}_{\tau_*}^2 \rangle < \langle \Delta \mathcal{Q}_{\tau_{\times}}^2 \rangle < (k_B T)^2$ (gray area), III) non-monotonic increase with at least one global maximum followed by a local minimum at $\tau_{\times}$ and $\tau_*$, respectively, such that  $ \langle \Delta \mathcal{Q}_{\tau_{*}}^2 \rangle <  \langle \Delta \mathcal{Q}_{\tau_{\times}}^2 \rangle = (k_B T)^2$ (pale blue area). The plots in the middle and right panels illustrate such distinct behaviors as a function of $\tau$ of the VACF and the variance of the exchanged energy, respectively, for specific values of the parameters $\eta_0 / \eta_{\infty}$ and $\lambda$, which are marked by the following symbols in the main diagram: $\eta_0 / \eta_{\infty} = 16$, $\lambda = 0.01 \tau_c$  ($\bigcirc$); $\eta_0 / \eta_{\infty} = 16$, $\lambda = 0.1 \tau_c$  ($\square$); and $\eta_0 / \eta_{\infty} = 16$, $\lambda = \tau_c$  ($\triangle$). The particular values $\tau_{\times}$ and $\tau_*$ of $\tau$ that give rise to the maxima and minima of $\langle \Delta \mathcal{Q}_{\tau}^2 \rangle$ are represented by the symbols $\times$ and $*$, respectively. In all cases, the variance of the energy exchange reaches the steady-state value $\langle \Delta \mathcal{Q}_{\tau \rightarrow \infty}^2 \rangle = (k_B T)^2$ set by the bath temperature $T$.}\label{fig:6}
\end{figure}

The previous analysis suggests that, unlike Brownian motion in a Newtonian liquid, in presence of the Basset-Boussinesq force distinct behaviors of the variance of the energy transfer in a viscoelastic liquid as a function of the time span $\tau$ are possible depending on the specific values of the dimensionless parameters characterizing the viscoelasticity of the medium, namely, $\eta_0 / \eta_{\infty}$ and $\lambda / \tau_c$. This is illustrated in the diagram of Fig. \ref{fig:6}, which discloses the existence of three types of behaviors of $\langle \Delta \mathcal{Q}_{\tau}^2 \rangle$ over 2 orders of magnitude of the viscosity ratio, $1 \le \eta_0 / \eta_{\infty} \le 128$, and 5 orders of magnitude for the stress relaxation time, $10^{-2} \le \lambda / \tau_c \le 10^3$, which can be identified as follows:

\begin{itemize}
	\item[I)]{Monotonic increase of $\langle \Delta \mathcal{Q}_{\tau}^2 \rangle$ as a function of $\tau$ (black area in Fig. \ref{fig:6}). In this regime the elasticity is so weak or the elastic stress relaxes so fast that the liquid cannot store a significant amount of energy during the exchange processes with the particle, thus effectively behaving as in a Newtonian liquid with a viscosity higher than that of the solvent. In this regime, the VACF is a monotonic decaying function of $\tau$ with no anticorrelations.}
	\item[II)]{Non-monotonic increase of $\langle \Delta \mathcal{Q}_{\tau}^2 \rangle$ as a function of $\tau$ with a local maximum at $\tau = \tau_{\times}$ followed by a local minimum at $\tau = \tau_* > \tau_{\times}$, such that $\langle \Delta \mathcal{Q}_{\tau_*}^2 \rangle < \langle \Delta \mathcal{Q}_{\tau_{\times}}^2 \rangle < (k_B T)^2$ (gray area in Fig. \ref{fig:6}). In this regime, the transient storage of energy by the liquid is noticeable for values of $\tau$ around $\tau_*$, which coincides with the local maximum of the VACF, whereas the location of $\tau_{\times}$ corresponds to the local minimum of the VACF.  However, the elasticity of the liquid is not sufficient to induce negative correlations of the particle velocity.}
	\item[III)]{Non-monotonic increase of $\langle \Delta \mathcal{Q}_{\tau}^2 \rangle$ as a function of $\tau$ with at least one global maximum at $\tau = \tau_{\times}$ followed by a local minimum at $\tau = \tau_* > \tau_{\times}$, such that  $ \langle \Delta \mathcal{Q}_{\tau_{*}}^2 \rangle <  \langle \Delta \mathcal{Q}_{\tau_{\times}}^2 \rangle = (k_B T)^2$ (pale blue area in Fig. \ref{fig:6}). In this regime, the transient storage of energy by the liquid is significant for values of $\tau$ around $\tau_*$, which coincides with the global minimum of the VACF around which anticorrelations of the particle velocity occur due to the sufficiently strong elasticity of the liquid, while the location of $\tau_{\times}$ coincides with the finite value of $\tau$ at which $\langle v(\tau_{\times}) v(0) \rangle_{eq} = 0$, right before becoming negative and reaching the global minimum at $\tau = \tau_*$.}
\end{itemize}

\section{Summary and concluding remarks}\label{sect:summ}

In this paper, we have examined in detail the statistical properties of the energy that is stochastically transferred between a spherical Brownian particle and its surrounding liquid medium at constant temperature in absence of externally applied forces, which gives rise to temporal variations of its kinetic energy. Based on the generalized Langevin equation that describes the underdamped particle motion in a liquid of arbitrary relaxation modulus including the Basset-Boussinesq force, we have derived an analytic expression for the probability density function of the energy exchanged between the particle and its liquid surroundings over a finite time interval. We have demonstrated that this probability distribution can be written as a zeroth order modified Bessel function of the second kind, whose corresponding variance can be explicitly expressed in terms of the velocity autocorrelation function of the particle, thereby including memory effects resulting from the hydrodynamic backflow and the viscoelasticity of the liquid. By means of numerical simulations of the non-Markovian dynamics of a particle in a Newtonian liquid including hydrodynamic memory, we have verified the derived expressions for representative values of the parameters of the system, thus validating our theoretical approach over many orders of magnitude of the time span over which the energy is exchanged. On this basis, in addition to hydrodynamic memory we have also analyzed the effect of the viscoelasticity of the surrounding liquid on the behavior of the variance of the exchanged energy, which have allowed us to unveil the existence of three distinct regimes of the stochastic energy transfer depending on the parameters characterizing the viscoelasticity of the liquid. In all situations, our general findings reveal that memory effects due to hydrodynamic backflow and viscoelasticity amplify the energy exchange process with respect to that taking place in a Markovian viscous bath on time intervals shorter than the momentum relaxation time of the embedded particle.

In recent years there has been an upsurge of interest in understanding stochastic thermodynamic quantities in mesoscopic systems owing to the possibility of utilizing micron- and submicron-sized systems as small-scaled devices, where most efforts have focused on models that usually neglect memory effects that naturally arise in their dynamics due to their coupling with the surroundings. Therefore, the results presented here provide extensions and generalizations beyond the paradigm of Brownian motion in idealized Markovian baths, thus shedding light on the impact of hydrodynamic and viscoelastic memory effects on the thermal energy transfer taking place in mesoscopic systems embedded in simple and complex fluids at short time scales.  Investigating 
the stochastic energetics and thermodynamics of small systems subject to hydrodynamic backflow and viscoelasticity of the environment in presence of well-controlled external forces, flows and gradients. as commonly encountered in experiments, could also be interesting for practical applications and will be studied elsewhere.

\section*{Acknowledgments}
We acknowledge support from DGAPA-UNAM PAPIIT Grant No. IA104922.

\appendix
\section{Simulation of stochastic dynamics including the Basset-Boussinesq force}\label{app:simul}

In this Appendix, we provide more details on the numerical simulation of trajectories evolving according to the generalized Langevin equation (\ref{eq:GLENewton}) that includes the Basset-Boussinesq force as a coarse-grained model for the stochastic motion of a spherical particle in a Newtonian liquid. We employ a Markovian embedding method \cite{Siegle2010,Siegle2011}, which allows one to recast Eq. (\ref{eq:GLENewton}) as a system of equations that are local in time by introducing a finite number of auxiliary variables. Following this approach, for $ t \ge t'$ the memory kernel defined through Eq. (\ref{eq:GSFour}) can be written as
{\footnotesize{
\begin{eqnarray}\label{eq:approxkernel}
	\fl \Gamma(t - t')  =  2\gamma \delta(t - t')  
			  +   \frac{\gamma C_{3/2}(b)}{2}\sqrt{\frac{\tau_{\mathrm{f}}}{\pi}} \left\{ 2\sum_{i = 1}^{n}\left( \frac{\nu_0}{b^i} \right)^{1/2} \delta(t -  t')  -   \sum_{i = 1}^{n} \left( \frac{\nu_0}{b^i} \right)^{3/2} \exp \left[- \frac{\nu_0}{b^i} \left( t -t' \right) \right] \right\},
\end{eqnarray}}}
and $\Gamma(t - t') = 0$ otherwise, where $\gamma = 6\pi a \eta$ is the steady-state friction coefficient of the particle immersed in the liquid of constant viscosity $\eta$, so that the first term on the right-hand side is related to the Stokes force acting on it. On the other hand, the second term on the right-hand side of Eq. (\ref{eq:approxkernel}) is associated to the Basset-Boussinesq force, where the instantaneous term accounts  for the singularity at $t' = 0$, while the weighted sum of $n$ exponentially decaying functions represent a fractal-scaling approximation of the resulting power-law behavior with exponent $-3/2$ of the kernel convoluted with the particle velocity. In Eq. (\ref{eq:approxkernel}). $\nu_0$ is a high cutoff frequency in the approximation, whereas $b > 1$ is a scaling dilation parameter that determines the low cutoff frequency $\nu_0 b^{-n}$, in such a way that the approximation is valid over approximately $n \log_{10}b  -2$ decades in the time interval $\nu_0^{-1}< t - t' < b^n \nu_0^{-1}$. In addition, $C_{3/2}(b)$ is a prefactor that depends on the particular choice of the value of $b$, where the subindex $3/2$ highlights the asymptotic power-law behavior of the kernel. Then, the following system of $n+2$ coupled equations including $n$ auxliary stochastic variables , $u^{(1)}(t), \ldots, u^{(n)}(t)$, can be introduced in order to mimic the non-Markovian dynamics using the approximation of the memory kernel given in Eq. (\ref{eq:approxkernel})
\begin{eqnarray}\label{eq:systSDE}
	\dot{x}(t) & = & v(t), \nonumber\\
	M \dot{v}(t) & = & - \left( \gamma + \gamma_0 \right) v(t) - \sum_{i = 1}^n u^{(i)}(t) + \sqrt{2 k_B T} \left[\sqrt{\gamma_0}\xi_0(t) +  \sqrt{\gamma} \xi_{n+1}(t) \right],\nonumber\\
\dot{u}^{(i)}(t) & = & - \nu_i u^{(i)}(t) - \gamma_i v(t) + \sqrt{ 2 k_B T \gamma_i \nu_i} \xi_i(t), \,\,\,\,\, i = 1, \ldots , n,
\end{eqnarray}
where $\nu_i = \nu_0 b^{-i}$, $\gamma_0 = \frac{1}{2}\gamma C_{3/2}(b) \sqrt{\frac{\tau_{\mathrm{f}}}{\pi}} \sum_{i = 1}^n \sqrt{\nu_i}$, $\gamma_i =  \frac{1}{2}\gamma C_{3/2}(b) \sqrt{\frac{\tau_{\mathrm{f}}}{\pi}} \nu_i^{3/2}$, $\xi_j(t)$ ($j = 1,\ldots , n+1$), are independent Gaussian white noises of zero mean and delta-correlated, i.e. $\langle \xi_j(t) \rangle = 0$ and $\langle \xi_j(t)  \xi_l(t') \rangle = \delta_{jl} \delta(t-t')$ with $l = 1,\ldots , n+1$ and $\delta_{jl}$ the Kronecker delta, and $\xi_0(t) = \sum_{i = 1}^n \sqrt{\frac{\gamma_i}{\nu_i \gamma_0}} \xi_i(t)$. Upon rescaling all the relevant physical quantities in the system of stochastic differential Eqs. (\ref{eq:systSDE}) by the characteristic scales defined in Eqs. (\ref{eq:tauc})-(\ref{eq:Fc}), i.e. $t \rightarrow t/ \tau_c$, $x \rightarrow x/x_c$, $v \rightarrow v/v_c$, $u^{(i)} \rightarrow u^{(i)}/F_c$, $\nu_0 \rightarrow \nu_0 \tau_c $,  $\nu_i \rightarrow \nu_i \tau_c $, $\tau_{\mathrm{f}} \rightarrow \tau_{\mathrm{f}} /\tau_c$, and $\gamma_i \rightarrow \gamma_i x_c/F_c$, along the discrete time $t_k = k \Delta t$ ($k = 0, 1, \ldots$) with constant time-step $\Delta t$, its discretized version using the Euler-Maruyama method reads
\begin{eqnarray}\label{eq:systSDEdisc}
    x_{k+1} & = & x_k + \Delta t v_k, \nonumber\\
    v_{k+1}& = & \left[ 1 - \left( 1 + \delta \right) \Delta t \right] v_k  - \Delta t \sum_{i = 1}^n  u_k^{(i)} + \sqrt{2 \Delta t} \left( \sqrt{\delta} N_0 + N_{n+1} \right), \nonumber\\
    u_{k+1}^{(i)} & = & \left( 1 - \nu_i \Delta t \right) u_k^{(i)} - \gamma_i \Delta t  v_k + \sqrt{2 \gamma_i \nu_i \Delta t} N_i, \,\,\,\,\, i = 1, \ldots , n.
\end{eqnarray}
In Eqs. (\ref{eq:systSDEdisc}) $\delta = \gamma_0 / \gamma$, $N_j $ ($j = 1, \ldots, n+1$) are independent random number drawn from the standard normal distribution at each step, and $N_0 = \sum_{i = 1}^n \sqrt{\frac{\gamma_i}{\nu_i \delta }} N_i$. We choose $n = 13$, $b = 5$, $C_{3/2}(b) = 1.78167$, and $\nu_0 = 10^3$, whereas the time-step is selected as $\Delta t = 10^{-4} < \nu_0^{-1}$ to achieve numerical stability of the solutions, for which the approximation in Eq. (\ref{eq:approxkernel}) is excellent over $n \log_{10}b -2 \approx 7$ decades in time. For the sake of simplicity, we set $x_0 = 0$ as the initial condition for the particle position. Moreover, to guarantee that the particle velocity is at all times in equilibrium with the bath at constant temperature, the initial condition $v_0$ must be drawn from the standard normal distribution. Finally, to ensure the stationarity at all times of the thermal noise $\zeta(t)$ in Eq. (\ref{eq:GLENewton}), whose statistical properties are correctly simulated by means of the auxiliary variables $u^{(i)}$ ($i = 1,\ldots, n$) and the random numbers $N_j$ ($j = 0, \ldots, n+1$), the initial conditions $u^{(i)}_0$ are required to be independent Gaussian random numbers with zero mean and variance $\gamma_i$.

\end{document}